\documentclass[twoside,11pt]{article}
\usepackage{graphicx}
\graphicspath{ {./images/} }
\usepackage{bm}
\usepackage{float}
\usepackage{physics}
\usepackage[round]{natbib}
\usepackage{tabularx}
\usepackage{multirow,multicol, makecell}
\usepackage{booktabs}
\usepackage{lipsum}

\usepackage{natbib}

\usepackage{appendix}
\usepackage[margin=1in]{geometry}
\usepackage[usenames]{color}
\usepackage{amsfonts, amssymb, amsmath, amsthm, multicol}
\usepackage[colorlinks=true, pdfstartview=FitV, linkcolor=blue,
citecolor=blue, urlcolor=blue]{hyperref}
\usepackage[usenames]{color}
\definecolor{Red}{rgb}{1,0.05,0}
\definecolor{Grn}{rgb}{0.1,0.7,0.1}
\definecolor{Blu}{rgb}{0.1,0.1,0.6}
\definecolor{Org}{rgb}{1,0.45,0}
\definecolor{Vio}{rgb}{0.6578,0,0.9478}
\definecolor{Mag}{rgb}{1,0.2,0.3}
\usepackage{authblk}
\usepackage{booktabs}
\usepackage{breqn}

\usepackage[section]{placeins}
\usepackage{graphicx}
\usepackage{amssymb}
\usepackage{epstopdf, epsfig}
\usepackage{color}
\usepackage{float}
\usepackage{amsmath}
\usepackage{tabularx}
\usepackage{bm}
\usepackage{accents}
\usepackage{natbib}
\usepackage{array}
\usepackage{tabularx}
\usepackage{multirow,multicol, makecell}
\usepackage{pdflscape}
\usepackage{float}
\usepackage{breqn}
\usepackage{mathtools}
\usepackage{breqn}
\usepackage[flushleft]{threeparttable}
\usepackage{booktabs,caption}
\usepackage{footnote}
\usepackage{amssymb}
\usepackage{rotating}
\usepackage{lscape}
\makesavenoteenv{tabular}
\newcolumntype{C}[1]{>{\centering\arraybackslash}p{#1}}
\newcolumntype{L}[1]{>{\raggedright\arraybackslash}p{#1}}
\newcolumntype{M}[1]{>{\centering\arraybackslash}m{#1}}
\usepackage[mathlines,displaymath]{lineno}
\usepackage{lscape}

\usepackage{mathtools}

\newcommand{\circledtext}[1]{\raisebox{.5pt}{\textcircled{\raisebox{-.9pt} {#1}}}}

\usepackage{hyperref}
\hypersetup{
	colorlinks,
	linkcolor={red!50!black},
	citecolor={blue!50!black},
	urlcolor={blue!80!black}
}

\usepackage{adjustbox}
\usepackage{xcolor}
\usepackage{soul}

\newtheorem{rmk}{Remark}
\newtheorem{nt}{Note}

\numberwithin{rmk}{section}
\numberwithin{nt}{section}

\title{Revisiting \citet{liu2006instantaneous} and \citet{zigunov2024one}: on the Equivalence of the Omnidirectional Integration and the Pressure Poisson Equation}

\author[1]{\textcolor{Blu}{Connor Pryce}}
\author[1]{\textcolor{Vio}{Lanyu Li}}
\author[1]{\textcolor{Red}{Zhao Pan}\thanks{To whom correspondence may be addressed: zhao.pan@uwaterloo.ca}}
\affil[1]{University of Waterloo, Dept. of Mechanical and Mechatronics Engineering, Waterloo, ON, Canada}

\date{\today}

\begin{document}
	\maketitle \vspace{1cm}

	\begin{abstract}
		\normalsize
		In this work, we demonstrate the equivalency of the Rotating Parallel Ray Omnidirectional Integration (RPR-ODI) and the Pressure Poisson Equation (PPE) for pressure field reconstruction from corrupted image velocimetry data (dubbed `$\text{ODI} \equiv \text{PPE}$').
		Building on the work by \citet{zigunov2024one}, we show that performing the ODI is equivalent to pursuing the \textcolor{black}{minimum norm least squares (MNLS) or minimum norm (MN) solution to a Poisson equation with all Neumann boundary conditions.} 
		By looking through the lens of linear algebra,  regression, optimization, and the well-
		posedness of the Poisson equation, we provide a comprehensive and integrated framework to analyze ODI/PPE-based pressure field reconstruction methods. 
		\textcolor{black}{The new comprehensions on $\text{ODI} \equiv \text{PPE}$ provides theoretical and computational insights valuable to experimentalists beyond reducing the \textcolor{black}{high} computational cost of ODI to that of PPE. More importantly, we i) provide  a comprehensive guideline for robust pressure reconstruction, and ii) unveil the shared strengths and limitations of ODI and PPE, which are elaborated in remarks and notes throughout this work.}
		Some remarks suggest simple regularization strategies that serve as `minimal reproducible examples' and provide a foundation for further refinement.
		This work paves the way for further improvements in ODI/PPE-based pressure field reconstruction by utilizing the extensive literature on fast and robust elliptic solvers as well as their associated regularization methods.
		Numerical experiments are presented to support and illustrate these arguments. 
		
	\end{abstract}

	\section{Introduction}
	
	Pressure field reconstruction from noisy velocity data obtained via Particle Image Velocimetry (PIV) or Lagrangian Particle Tracking (LPT) is an effective and non-invasive flow diagnostic strategy \citep{van2013piv}. 
	Since early development by \citet{schwabe1935druckermittlung}, continuous progress in the field has resulted in a wide range of viable reconstruction methods. 
	These reconstruction methods typically fall into one of two categories. 
	The first category {reconstruct} the pressure field $p$ by integrating the pressure gradient, $\bm{g}(\bm{u})$, obtained using the momentum equation
	\begin{equation}
		\label{eq: grad p}
		\nabla p = \bm g(\bm{u}) = -{\rho} \left[\frac{\partial   \bm{u}}{\partial t} + (\bm{u} \cdot \nabla)\bm{u} - \nu\nabla^2 \bm{u}\right] \quad \text{in}~ \Omega, 
	\end{equation}
	where $\bm{u}$ is the velocity field, $\rho$ is the density, and $\nu$ is the kinematic viscosity, \textcolor{black}{for an incompressible flow.} 
	
	One iconic method believed to be in this category, Omni-Directional Integration (ODI), was first introduced by \citet{liu2003measurements,liu2006instantaneous}, with the key idea of trying to enforce the path independence of the pressure gradient by iteratively averaging the integrated pressure along different paths starting from a virtual boundary that encloses the domain.
	This reconstruction method is known to be particularly robust to random noise in the pressure gradients, however, due to the ensemble practice, the original ODI \textcolor{black}{methods can be very} expensive \citep{zigunov2024one}. To improve the computational efficiency, several strategies have been implemented. 
	Examples include using much fewer integral paths \citep{dabiri2013algorithm}, GPU-acceleration \citep{wang2019gpu}, Green's functions \citep{wang2023greens}, and removing redundant integration \citep{zigunov2023fast,zigunov2024one}. 
	
	In addition, the reconstruction quality of ODI depends on the size and shape of the virtual boundary. 
	Since its inception, the virtual boundary has evolved from finite rectangular virtual boundary \citep{liu2003measurements}, to finite circular virtual boundary \citep{liu2006instantaneous}, and finally to an infinitely large virtual boundary which is the Rotating Parallel Ray Omni-Directional Integration (RPR-ODI) \citep{liu2016instantaneous,wang2019gpu}. 
	Due to the unbiased weighting of the ensemble integration, RPR-ODI is in general considered superior in terms of reconstruction accuracy among all other conventional ODI solvers. 
	
	The second category of methods recovers the pressure field using the Pressure Poisson Equation (PPE) which is derived by applying the divergence to both sides of \eqref{eq: grad p}:
	\begin{equation}
		\label{eq: PPE ideal}
		\nabla ^2 p  = f(\bm{u}) = \nabla \cdot \bm g(\bm{u}) = -{\rho} \nabla \cdot  \left[\frac{\partial   \bm{u}}{\partial t} + (\bm{u} \cdot \nabla)\bm{u} - \nu\nabla^2 \bm{u} \right] \quad \text{in}~\Omega,  
	\end{equation}
	where $f(\bm{u}) = \nabla \cdot \bm{g}(\bm{u})$ is the data of the Poisson equation.
	Invoking the vector calculus identity $\nabla\cdot(\nabla^2 \bm{u}) = \nabla^2  (\nabla\cdot\bm{u})$, the viscous term in \eqref{eq: PPE ideal} vanishes if the flow is incompressible. 
	Although the Poisson equation benefits from superior numerical stability, ease of implementation, and high computational efficiency, experimentation using various numerical methods \citep{charonko2010assessment,sperotto2022meshless,zhang2020using}  has shown that reconstruction may suffer, more or less, when subject to noisy velocimetry data.
	
	Over the years, it has long been debated which of the two categories is better for pressure reconstruction, particularly the ODI versus PPE. 
	Some have claimed that the ODI is fundamentally different and far more accurate than solving the PPE (e.g., \citet{liu2020error}); while other studies have observed similar reconstruction accuracy between the ODI and PPE \citep{mcclure2017optimization}. 
	This debate has been filled with confusion and conflicting results until a recent breakthrough by \citet{zigunov2023fast,zigunov2024one} while trying to improve the computational efficiency of ODI. 
	In a series of works, \citet{zigunov2023fast,zigunov2024one} reformulated the iterative integration process of the ODI into a system of linear equations resembling the discretized PPE, alluding to a deep connection between the ODI and PPE.

	Inspired by the recent progress of \citet{zigunov2023fast,zigunov2024one}, in this work, we show that indeed the RPR-ODI is equivalent to a special case of PPE, and that the reason for the perceived difference in robustness between ODI and PPE is all in the details of numerical implementation and regularization. 
	These findings will end a long debate about which method is better, and reduce the computational cost of the ODI to match that of the PPE while preserving the robustness observed in ODI. Additionally, and more importantly, we establish a unified framework to analyze ODI/PPE-based pressure field reconstruction through the lenses of the well-posedness of the elliptic Poisson equation, linear algebra, estimation, and optimization.  
	While the tools used in this work are well-established, we aim to provide an integrated analysis offering comprehensive insights and hopefully inspire new, effective regularization techniques for faster, more accurate, and robust pressure reconstruction.
	
	The work is organized as follows. 
	Sect.~\ref{sec: ODI = PPE} establishes the equivalency between the ODI and PPE by drawing on the results in \citet{zigunov2024one}. Sect.~\ref{sec: Well-posedness of the Neumann problem} reviews the major results of the well-posedness of the Poisson equation in the continuous and discrete settings, which serves as a ground for further analysis.
	In Sect.~\ref{sec: underdetermined regu}, we unveil the precise meaning of the solution from the ODI.
	Sect.~\ref{sec: robustness of ODI/PPE} gives insights into the robustness of the ODI/PPE-based pressure field reconstruction, followed by Sect.~\ref{sec: one point DBC}, where the singularity of the one-point Dirichlet setup for a PPE is discussed.
	\textcolor{black}{We also provide a comprehensive guideline for robust pressure field reconstruction with notes for practitioners in Sect.~\ref{sec: Computational notes}, before demonstrating synthetic experiments supporting our major arguments throughout the work in Sect.~\ref{sec: validation}.}
	We conclude this work by recapping the shared fundamentals of the ODI and PPE in Sect.~\ref{sec: conclusion}, which also infer perspectives for future improvements.
	
	
	

	\section{Equivalency of ODI and PPE }
	\label{sec: ODI = PPE}

	In this section, we demonstrate that using the RPR-ODI to reconstruct the pressure field is equivalent to pursuing a Minimal Norm (MN) or Minimal Norm Least Squares (MNLS) solution to a PPE with all Neumann boundary conditions.
	
	The recent advancement of ODI \citep{zigunov2023fast} removed the redundant ensemble integration along different paths by examining the integral at each cell in the domain and on the boundaries. 
	The resulting iterative pressure solver named ``Iterative Matrix ODI (I-MODI)'' by the authors is significantly faster than traditional versions of ODI.
	This work showed that at convergence, the accuracy of the I-MODI solver is typically very close to or slightly better than the {Rotating} Parallel Ray ODI (RPR-ODI), which is considered the most accurate among all of the traditional ODI solvers \citep{liu2016instantaneous,wang2019gpu}.
	More importantly, it was shown that when given the same corrupted velocimetry data, the reconstruction \textit{error} from the I-MODI and high-resolution RPR-ODI are almost identical.
	This is strong evidence suggesting that the matrix ODI approaches the limit of the traditional ODI when computed with high resolution.

	A follow-up work \citep{zigunov2024one} examined the fixed point of the I-MODI iteration and found that at convergence, ODI can be computed by solving a linear system of the form
	\begin{equation}
		\label{eq: linear system}
		\bm{\mathcal{L}} {\vb{p}} = {\vb{b}},
	\end{equation}
	with ${\vb{p}}$ being the pressure field to be solved, ${\vb{b}}$ the data, and $\bm{\mathcal{L}}$ is the discretized Laplacian.\footnote{
		\textcolor{black}{In this work, we use variables with upright bold font to indicate vectors in the context of linear algebra (e.g., $\vb{p}$), and italic variables for continuous setting (e.g., $p$ is a scalar field and $\bm{g}$ and $\bm{u}$ are vector fields).}}
	This linear system does not have to be solved iteratively and is dubbed as ``One-Shot Matrix ODI'' (OS-MODI).
	When solved for the pressure field, it was shown that OS-MODI and I-MODI result in the same pressure reconstruction as RPR-ODI, with an $O(10^3)$ to $O(10^6)$ times improvement in speed on their test cases.

	Beyond the robustness and high computational efficiency, the reformulation in \citet{zigunov2024one} suggests strong connections between PPE and ODI through the numerical implementation of $\bm{\mathcal{L}}$ and $\vb{b}$ in \eqref{eq: linear system}. 
	\textcolor{black}{In the following sections, we will use different approaches to demonstrate the equivalency between the ODI methods (particularly the OS-MODI) and PPE on an irregular and regular mesh.}
	
	\subsection{Interpretation by finite volume method}
	\label{sec: FVM derivation}
	
	\color{black}
	This section provides an alternative derivation of the stencils for OS-MODI developed by \citet{zigunov2024one} using the finite volume method.
	
	The Poisson equation \eqref{eq: PPE ideal} equipped with all Neumann boundary conditions is 
	\color{black}
	\begin{equation} 
		\label{eq: PPE with NBC}
		\begin{split}
			\nabla^2 p & = f = \nabla \cdot \bm{g} ~ \quad \text{in} \quad \Omega \\
			\bm{n} \cdot \nabla p & = g_n = \bm{n} \cdot \bm{g} \quad \text{on} \quad \partial\Omega, 
		\end{split}
	\end{equation}
	\color{black}
	where $f$ and $g_n$ are the data in the domain and on the boundaries, respectively. 
	Multiplying by a weighted test function $\omega_h$ and integrating over the domain provide a weak formulation of the Poisson equation. Rearranging leads to 
	\begin{equation}
		\label{eq: weak PPE}
		\int_{\Omega} \nabla\omega_h \cdot \nabla p dV  - \int_{\Omega} \nabla\omega_h \cdot \bm{g} dV  =  \int_{\partial\Omega} \omega_h ( \nabla p -  \bm{g}) \cdot \bm{n} dS = 0.
	\end{equation}
	The second equal sign in \eqref{eq: weak PPE} holds when the Neumann boundary condition is enforced.
	
\begin{figure}
		\centering
		\includegraphics[width=0.95\textwidth]{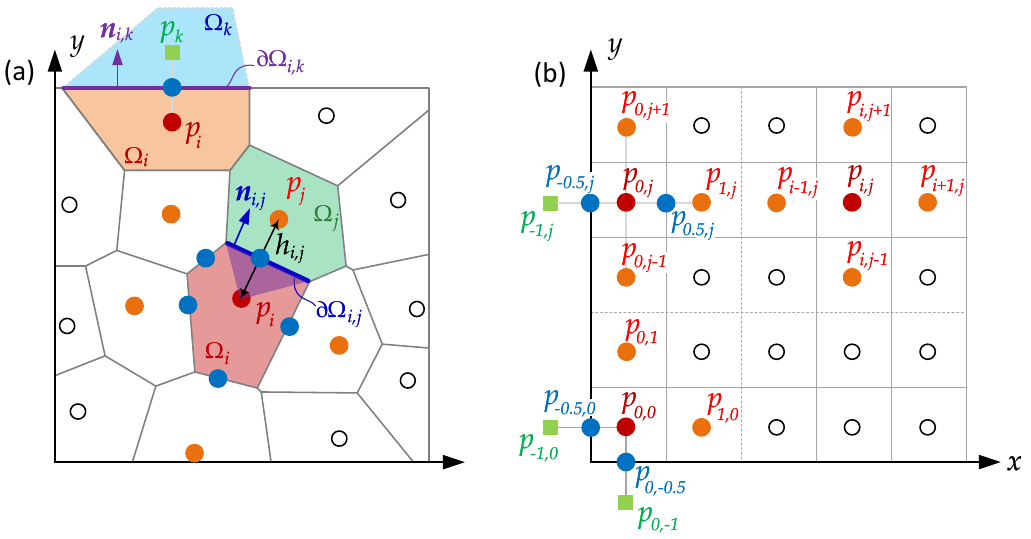}
		\caption{Mesh and discretization on irregular and regular data. (a) Voronoi diagram for a domain with scattered points. 
			The pressure value at the central Voronoi cell ($\Omega_i$, shaded in red) is $p_i$, and $p_j$ is the pressure at the node of one of the neighboring cell ($\Omega_j$, shaded in green). The interface between $\Omega_i$ and $\Omega_j$ (thick blue line) is denoted as $\partial\Omega_{i,j}$. The blue shaded zone indicates a ghost cell ($\Omega_{k}$) outside the domain $\Omega$.  (b)~Cell-centered layout of the discretized domain on a regular grid. The green squares indicate the ghost nodes.
		}
		\label{fig: voroni}
	\end{figure}

	Next, we use the finite volume approximation to discretize \eqref{eq: weak PPE} on a Voronoi mesh.
	Given the measurement of the pressure gradients ($\bm{g}_i$) available at scattered node $\bm{x}_i$ (indicated by the circles, filled or open, as shown in Fig.~\ref{fig: voroni}(a)), we discretize the domain $\Omega$ by Voronoi tessellation, and the control volume for $\bm{x}_i$ is $\Omega_i$.
	The pressure gradient at the interface between the central cell $\Omega_i$ and a neighboring cell $\Omega_j$  can be approximated as 
	\begin{equation}
		\label{eq: grad p num}
		\nabla p_{i,j} = \frac{p_j - p_i}{h_{i,j}} \bm{n}_{i,j},
	\end{equation}
	where $h_{i,j}$ is the distance between the two nodes, and $\bm{n}_{i,j}$ is the unit normal vector pointing from $\bm{x}_i$ towards the neighboring node $\bm{x}_j$. 
	
	Using the finite volume method for cell $\Omega_i$ to approximate the first term in \eqref{eq: weak PPE} and invoking \eqref{eq: grad p num} leads to
	\begin{equation}
		\label{eq: FVM for grap p}
		\int_{\Omega_i} \nabla\omega_h \cdot \nabla p dV = \sum_{j} \int_{\partial\Omega_{i,j}} \omega_h \nabla p \cdot \bm{n}_{i,j} dS \approx \sum_{j} {\omega_h}_{i,j} |\partial \Omega_{i}| \frac{p_j - p_i}{h_{i,j}},
	\end{equation}
	where $\partial\Omega_{i,j}$ is the interface of the neighboring cells $\Omega_{i,j}$, $|\partial\Omega_i|_j$ is the length (or area) of $ \partial\Omega_{i,j}$ depending on the dimension of the domain. 
	
	Invoking the property of the Voronoi diagram---where the interface lies halfway between adjacent nodes\footnote{This implies that the success of the OS-MODI on irregular data as demonstrated in \citet{zigunov2023fast,zigunov2024one} also partially depends on the use of Voronoi tessellation.}, the pressure gradients at the interface can be approximated by averaging the pressure gradients measured at the adjacent nodes
	\begin{equation}
		\label{eq: g = 0.5(g+g)}
		\bm{g}_{i,j} =  \frac{\bm{g}_i + \bm{g}_j}{2}.
	\end{equation}
	Similar to \eqref{eq: FVM for grap p}, the second term in \eqref{eq: weak PPE} can be approximated as 
	\begin{equation}
		\label{eq: FVM for g}
		\int_{\Omega_i} \nabla\omega_h \cdot \bm{g} dV = \sum_{j} \int_{\partial\Omega_{i,j}} \omega_h \bm{g} \cdot \bm{n}_{i,j} dS \approx \sum_{j} {\omega_h}_{i,j} |\partial \Omega_{i}| \bm{g}_{i,j} \cdot \bm{n}_{i,j},
	\end{equation}
	where $\bm{g}_{i,j}$ is the measured pressure gradient at the interface $\partial\Omega_{i,j}$.
	

	Substituting \eqref{eq: FVM for grap p} and \eqref{eq: FVM for g}  into \eqref{eq: weak PPE} and applying the Neumann condition lead to
	\begin{equation}
		\label{eq: FVM =}
		\sum_{j} {\omega_h}_{i,j} |\partial \Omega_{i}| \frac{p_j - p_i}{h_{i,j}}  = \sum_{j} {\omega_h}_{i,j} |\partial \Omega_{i}| \bm{g}_{i,j} \cdot \bm{n}_{i,j}.
	\end{equation}
	If we choose ${\omega_h}_{i,j} = h_{i,j}$ and invoke \eqref{eq: g = 0.5(g+g)}, \eqref{eq: FVM =} transforms to 
	\begin{equation}
		\label{eq: sum p-p = sum g+g}
		\sum_j |\partial\Omega_i|_j (p_j - p_i)  = \sum_j h_{i,j} |\partial\Omega_i|_j\frac{\bm{g}_i + \bm{g}_j}{2} \cdot \bm{n}_{i,j}.
	\end{equation}
	Noting that $|\partial\Omega_i| = \sum_j |\partial\Omega_i|_j$,  rearranging \eqref{eq: sum p-p = sum g+g} leads to 
	\begin{equation}
		\label{eq: pc = sum pj - g+g}
		p_i  = \sum_j \frac{|\partial\Omega_i|_j}{|\partial\Omega_i|}  \left(p_j - \frac{\bm{g}_i + \bm{g}_j}{2} \cdot \bm{n} h_{i,j} \right),
	\end{equation}
	which recovers the results by \citet{zigunov2024one}---particularly, equation (11) or (25) in their work.
	
	It is interesting to see that this choice of weight in \citet{zigunov2023fast} depends on the layout of the nodes. The weight applied to the pressure gradient is proportional to the separation between the nodes. A higher weight is given to the region where the data is sparse. 
	The numerical properties of this choice of weight are out of the scope of the current research, and we will leave this topic for future studies.
	
	When the data (and mesh) is isotropic, the Poisson equation can also be recovered directly from \eqref{eq: sum p-p = sum g+g}: 
	Setting $h_{i,j} = h$ to be a constant over the entire domain $\Omega$ 
	and dividing \eqref{eq: sum p-p = sum g+g} by $|\Omega_{i}|h$ leads to 
	\begin{equation}
		\label{eq: natural div}
		\sum_j \frac{|\partial\Omega_i|_j}{|\Omega_i|} \frac{p_j - p_i}{h}  = \sum_j \frac{|\partial\Omega_i|_j}{|\Omega_i|}\frac{\bm{g}_i + \bm{g}_j}{2} \bm{n}_{i,j} = \sum_j \frac{|\partial\Omega_i|_j}{|\Omega_i|}\bm{g}_{i,j} \bm{n}_{i,j},
	\end{equation}
	where $|\Omega_i|$ is the area (or volume) of $\Omega_i$ in 2D (or 3D).
	Recalling that the definition of the divergence of a vector field, say $\bm{g}$, is 
	$ \nabla \cdot \bm{g} = \lim_{\Omega \to 0} \frac{1}{|\Omega|} \int_{\partial\Omega} \bm{g}\cdot \bm{n} dS,$
	we can see that \eqref{eq: natural div} is a natural discretization \citep{hyman1997natural} for $\nabla\cdot\nabla p = \nabla \cdot \bm{g}.$
	In fact, if the mesh is uniform rectangular, \eqref{eq: FVM =} reduces to a stencil based on the cell-centered finite difference (see Fig.~\ref{fig: voroni}(b) for the node layout, and Sect.~\ref{sec: finite difference} for derivation).

	Here, we want to emphasize that consistent numerical computation of the data of the Poisson equation $f$ (e.g., $f = \nabla \cdot \bm{g}$ in \eqref{eq: PPE ideal}) is critical to ensure the compatibility condition and the existence of the solution for a Neumann problem. 
	We will soon elaborate on these aspects (e.g., see Sect.~\ref{sec: Well-posedness of the Neumann problem}, Remark~\ref{rmk: guaranteed compatibility} and Note~\ref{note: div therom}). 
	\textcolor{black}{For the stencils as in \eqref{eq: FVM =}, which allows the choice of arbitrary weight $\omega_h$, finite volume approximation preserves the `flux' (i.e., pressure gradient in our context) and ensures the compatibility of the corresponding Poisson equation. 
		This extends to the stencils developed by \citet{zigunov2024one}, which are subject to a specific choice of $\omega_h$.
		In other words, consistent computation of data is implicitly `hard-coded' through the use of finite volume approximation by applying the divergence theorem.}

	\subsection{Interpretation by finite difference on regular grid}
	\label{sec: finite difference}
	A special case of the general tessellation used in Sect.~\ref{sec: FVM derivation} is the mesh generated by regular squares (or cubes), which are particularly relevant for the data from PIV.
	In this section, we re-derive the results in \citet{zigunov2024one} on a 2D regular grid using finite difference methods.
	
	\color{black}
	\citet{zigunov2024one} showed that on a 2D uniform grid, at the convergence of ODI, the pressure at an interior nodal point ($p_{i,j}$) can be evaluated as
		\begin{multline} 	
			\label{eq: ODI stencil}
			p_{i,j} =  \frac{1}{4} \left(p_{i+1,j} + p_{i-1,j}  + p_{i,j+1} + p_{i,j-1} \right) \\
			+ \frac{1}{8}h\left(g_{x}|_{i-1,j} - g_{x}|_{i+1,j} - g_{y}|_{i,j+1} + g_{y}|_{i,j-1}\right),
		\end{multline}
		where $g_x|_{i,j}$ 
		and $g_y|_{i,j}$ are the pressure gradients in $x$ and $y$ directions obtained from the momentum equations using the measured velocity evaluated at location $(x_i,y_j)$, and $h$ is the grid spacing.\footnote{The notation for the pressure gradient at, for example, center and west nodal points are $f_{x}(C)$ and  $f_{x}(E)$, respectively, in \citet{zigunov2024one}; 
			while, in the current work, they are denoted as $g_{x}|_{i,j}$ and $g_{x}|_{i+1,j}$, respectively.} 
		Rearranging \eqref{eq: ODI stencil} leads to
		\begin{multline}
			\label{eq: Fernado interior nodal point}
			-4p_{i,j} +p_{i+1,j} + p_{i-1,j} + p_{i,j+1} + p_{i,j-1}  = \\ h^2 \left( \frac{g_{x}|_{i+1,j} - g_{x}|_{i-1,j}}{2h} + \frac{g_{y}|_{i,j+1} - g_{y}|_{i,j-1}}{2h}\right) = h^2f_{i,j},
		\end{multline}
		where the data $f_{i,j} = (\partial_x g_{x} + \partial_y g_{y})|_{i,j}$ is evaluated by the second order central difference (e.g., $\partial_x g_{x}|_{i,j} = \frac{1}{2h}(g_{x}|_{i+1,j} - g_{x}|_{i-1,j})$).  
		This result is the familiar five-point finite difference scheme for the discrete Laplacian~\citep{Chen2020lecture,thomas2013numerical_ellip,thomas2013numerical}, denoted as $(-4,1,1,1,1)$ where each value corresponds to the pressure coefficients seen in the left hand side of \eqref{eq: Fernado interior nodal point}. 
		The layout of the nodes and the stencil shown in \eqref{eq: Fernado interior nodal point} is illustrated in Fig.~\ref{fig: voroni}(b).
		%
		
		
		Likewise, \citet{zigunov2024one} also derived 
		\begin{multline}
			\label{eq: Fernado boudanry nodal point}
			p_{0,j} =  \frac{1}{3}\left(p_{1,j} + p_{0,j+1} + p_{0,j-1}\right)  \\ + \frac{1}{3}h \left( -\frac{g_{x}|_{0,j} + g_{x}|_{1,j}}{2} + \frac{g_{y}|_{0,j-1} - g_{y}|_{0,j+1}}{2}\right),
		\end{multline}
		which describes the treatment of the nodes on the boundary.
		To derive this stencil by finite differences, we start by introducing a ghost node on a cell-centered grid located a distance of $h/2$ away from the edge of the domain.\footnote{This ghost node based Neumann boundary condition implementation works beyond regular mesh. In Appendix~\ref{sec: generic interpolation}, we provide an alternative derivation of the stencil \eqref{eq: sum p-p = sum g+g} based on generic interpolation on an irregular mesh.}
		Then we enforce a Neumann condition on the boundary (e.g., specify the value of $\bm{n} \cdot \nabla p$ at $(x_{-0.5},y_j)$) using the following first-order scheme:
		\begin{equation}
			\label{eq: ghost point cell-centered 1}
			-\bm{n} \cdot \nabla p|_{-0.5,j} = g_{x}|_{-0.5,j} = \frac{  p_{0,j} - p_{-1,j} }{h}. 
		\end{equation}
		We compute $g_{x}|_{0.5,j}$ by linearly interpolating between $g_{x}|_{0,j} $ and $g_{x}|_{1,j}$:
		\begin{equation}
			\label{eq: ghost point cell-centered 2} 
			g_{x}|_{0.5,j} = \frac{g_{x}|_{0,j} + g_{x}|_{1,j}}{2}, 
		\end{equation}
		and evaluate the data $f_{0,j}$ by 
		\begin{equation}
			\label{eq: ghost point cell-centered 3} 
			f_{0,j} = \frac{ g_x|_{0.5,j} - g_x|_{-0.5,j}}{h} + \frac{g_y|_{0,j+1} - g_y|_{0,j-1}}{2h}.
		\end{equation}

		Substituting \eqref{eq: ghost point cell-centered 1} and \eqref{eq: ghost point cell-centered 2} into \eqref{eq: ghost point cell-centered 3}, and then \eqref{eq: ghost point cell-centered 3} into \eqref{eq: Fernado interior nodal point} and setting $i = 0$ recover \eqref{eq: Fernado boudanry nodal point}. 
		The stencil indicated in \eqref{eq: Fernado boudanry nodal point} can be denoted as $(-3,1,1,1)$ (see also in Fig.~\ref{fig: voroni}(b)). 
		A similar treatment using ghost points at the corners of the domain will produce stencil $(-2,1,1)$, which also matches \citet{zigunov2024one}, and is one of the classic implementations of boundaries and corners for a Neumann problem~\citep{Chen2020lecture,thomas2013numerical}.
		

		

		\textcolor{black}{
			The above combination of corner, wall, and interior stencils is a good choice for the Neumann problem of the Poisson equation; however, it is not the only one.}
		Employing other numerical methods 
		may also yield stencils similar to or better than the ones mentioned above. 
		In the current work, we focus on establishing the equivalency of ODI and PPE (dubbed as $\text{ODI} \equiv \text{PPE}$) using a simple finite difference scheme following \citet{zigunov2024one}. 
		
		Solving the resulting system of equations from ODI or the PPE with all Neumann conditions is non-trivial, as $\bm{\mathcal{L}}$, which we refer to as $\bm{\mathcal{L}}_N$ from now on with the subscript $[\cdot]_N$ to denote the Neumann problem, is rank-deficient, and thus, not invertible. 
		If a solution to \eqref{eq: linear system} exists, it is not unique; not to mention that a solution may not exist at all. 
		The numerical details in which this system is solved are important and will be elaborated upon in the following sections \textcolor{black}{after introducing the first major result of the current work:}  
		
		
		\begin{rmk}[ODI~$\equiv$~PPE] 
			\label{rmk: ODI = PPE}
			Using the ODI to reconstruct the pressure field from the pressure gradient is (numerically) equivalent to solving a PPE with all Neumann boundary conditions.
		\end{rmk} 
		\textcolor{black}{The chain of reasoning is straightforward, and we recap it here briefly: \citet{zigunov2023fast,zigunov2024one} developed the equivalency between the RPR-ODI and the OS-MODI, and we may dub this contribution as `$\text{RPR-ODI} \equiv \text{OS-MODI}$'. The current work showed that OS-MODI is equivalent to solving a Neumann problem of PPE (i.e., `OS-MODI~$\equiv$~PPE').
			Combining the above two equivalencies, we establish `RPR-ODI~$\equiv$~PPE', which is shorthanded as `ODI~$\equiv$~PPE'.}
		
		
		With the equivalence between the RPR-ODI and the PPE established, this work will now focus on exploring key theoretical aspects of ODI/PPE for pressure reconstruction addressing: 
		(i) what is the precise sense of the solution obtained from the ODI or equivalent PPE of~\eqref{eq: linear system} (see Sect.~\ref{sec: Well-posedness of the Neumann problem} and \ref{sec: underdetermined regu})? 
		(ii) Why is ODI/PPE robust against random noise, and why does it appear to perform particularly well with high-resolution data (see Sect.~\ref{sec: robustness of ODI/PPE})? 
		(iii) If {the} ODI and PPE are equivalent, why do some observe that the ODI demonstrates greater robustness compared to certain implementations of the PPE (see Sect.~\ref{sec: one point DBC})? 
		(iv) What insights can we gain from  ODI/PPE, and how can we make further improvements (see the notes and remarks scattered throughout the rest of the current work)?
		
		The discussions that follow primarily build on established results from the well-posedness of Poisson equations, linear algebra, estimation theory, and numerical analysis. 
		Together, they provide a unified framework for analyzing and refining ODI/PPE-based methods for pressure field reconstruction.

		\section{Well-posedness of the Neumann Problem of the Poisson Equation}
		\label{sec: Well-posedness of the Neumann problem}

		
		
		\textcolor{black}{
			In this section, we lay the foundation for the remainder of the current work by briefly reviewing the well-posedness of the Poisson equation in the continuous and discretized settings.
		}
		\textcolor{black}{
			The discretized form of \eqref{eq: PPE with NBC}  is expressed as
			\begin{equation}
				\label{discrete neumann problem}
				\bm{\mathcal{L}}_N \vb{p} = \vb{b}_N, 
			\end{equation}
			where $\bm{\mathcal{L}}_N$ emulates the Laplacian $\nabla^2$ with the Neumann boundary conditions, and $\vb{b}_N$ embeds the data $f$ and $g_n$ from its continuous counterpart.
			Upon proper discretization (e.g., as shown in Sect.~\ref{sec: ODI = PPE}), $\bm{\mathcal{L}}_N$ is symmetric\footnote{Although the Laplacian is self-adjoint, not necessarily all discretizations yield a symmetric $\bm{\mathcal{L}}_N$.} and semi-definite. 
		}
		
		\textcolor{black}{
			Both \eqref{eq: PPE with NBC} and \eqref{discrete neumann problem} are not well-posed.
			A problem is said to be well-posed if the following three conditions are satisfied: (i)~a solution exists, (ii)~the solution is unique, and (iii)~the solution continuously depends on the data. These conditions are summarized in the following sections and in Table~\ref{tab: ill-possed problem} with intuitions and remedies in the context of the current work.
		}
		
		\subsection{The Existence of a Solution}
		\label{sec: exist}
		
		The solution to \eqref{eq: PPE with NBC} exists if and only if the following compatibility condition is satisfied
		
		\begin{equation}
			\label{eq: compatibility condition clean}
			\int_{\Omega} f dV - \int_{\partial \Omega} g_n  dS = 0.
		\end{equation}
		This can be intuitively understood in the context of a steady-state diffusion problem with internal generation:
		The compatibility condition is satisfied if and only if the total flux (i.e., $\int_{\partial \Omega} g_n  dS$) through the boundary is balanced by the total generation in the domain (i.e., $\int_{\Omega} f dV$). 
		In many cases, if $f$ is not carefully calculated from $\bm{g}$, the compatibility condition \eqref{eq: compatibility condition clean} may not be satisfied.
		
		The compatibility condition can also be viewed in the discretized setting where the existence of a solution to \eqref{discrete neumann problem} requires that the data vector $\vb{b}_N$ lie in the column space of $\bm{\mathcal{L}}_N$ (i.e., $\vb{b}_N \in \text{Col}(\bm{\mathcal{L}}_N)$). Since $\vb{b}_N$ in \eqref{discrete neumann problem} embeds $f$ and $g_n$, it also must reflect the continuous compatibility condition of \eqref{eq: compatibility condition clean}.
		Here, the balance of total flux and generation in the domain manifests itself as zero-sum in the discretized setting (i.e., $\sum^m_{i=1} {b_N}_i = \vb{1}^\intercal \vb{b}_N = 0,$ where $m$ is the dimension of $\bm{\mathcal{L}}_N$, ${b_N}_i$ is the $i$-th element of ${\vb{b}_N}$, and $\vb{1} = (1,1,\dots,1)^\intercal$ is the one-vector). 
		This is equivalent to requiring that $\vb{b}_N$ be mean zero and thus orthogonal to~$\vb{1}$:
		\begin{equation}
			\label{eq: mean b = 0}
			\langle{\vb{b}_N}\rangle = \frac{1}{m} \sum^m_{i=1} {{b}_N}_i = \frac{1}{m} \vb{1}^\intercal \vb{b}_N =  0.
		\end{equation}
		This implies that the null space of $\bm{\mathcal{L}}_N^\intercal$ must be spanned by the one vector $\vb{1}$.
		If~\eqref{eq: mean b = 0} does not hold, $\vb{b}_N$ is incompatible, and the linear system is inconsistent.\footnote{Consistent and here, in the context of linear algebra, refers to a linear system that has a solution. In the context of pressure reconstruction based on the PPE, an inconsistent system is rooted in the incompatibility of the Neumann problem of the Poisson equation. In this work, we use consistency and compatibility interchangeably when discussing the existence of the solution in discretized or continuous settings.}
		Note that some non-standard discretizations may lead to a linear system that is compatible, however, the null space is not spanned by $\bm{1}$.

		\subsection{The Uniqueness of a Solution}
		\label{sec: uniq}
		
		If the solution to \eqref{eq: PPE with NBC} exists, it is unique up to an arbitrary constant. In other words, if $p$ solves \eqref{eq: PPE with NBC}, any $p + p_0$, where $p_0$ is a constant, is also a solution.
		This can be intuitively understood as pressure being a relative value defined with respect to a reference pressure.
		\textcolor{violet}{Common choices for the reference pressure can be the mean pressure in the domain or a pressure value at a certain location in the domain.}
		
		\textcolor{black}{In the discrete setting, the non-uniqueness is reflected by the fact that $\bm{\mathcal{L}}_N$ is not one-to-one and has a non-trivial null space. 
			Here, $\text{Null}(\bm{\mathcal{L}}_N)$ is also spanned by the one-vector $\vb{1}$, and this is not a coincidence since proper discretization guarantees a symmetric $\bm{\mathcal{L}}_N$. 
			As a result, the solution $\vb{p}$ is unique up to a scalar multiple of the one-vector $\vb{1}$ (i.e., $\vb{p} + \vb{p}_0,$ where $\vb{p}_0 = p_0\vb{1}$, is also a solution to \eqref{discrete neumann problem}).
			This recovers the heuristic in the continuous setting: $p$ is unique up to a constant $p_0$, not any other functions. 
			By symmetry, 
			$\text{Null}(\bm{\mathcal{L}}_N) = \text{Null}(\bm{\mathcal{L}}_N^\intercal)=\vb{1}$ and this implies that $\bm{\mathcal{L}}_N$ from a proper discretization should be rank-deficit by one and have one and only one eigenvalue.}
		
		\textcolor{black}{
			One way to enforce the uniqueness of \eqref{discrete neumann problem} is to remove the null space of $\bm{\mathcal{L}}_N$ by assigning a value to $\vb{1}^\intercal \vb{p}$, which is equivalent to prescribing the mean value for the pressure as any scalar ${p}_0$:
			\begin{equation}
				\label{eq: mean p = bar p0}
				\langle{\vb{p}}\rangle = \frac{1}{m}\vb{1}^\intercal \vb{p} = {p}_0.
			\end{equation}
		}
		
		\color{black}
		\subsection{Continuous Dependence}
		\label{sec: Continuous Dependence}
		If the solution to \eqref{eq: PPE with NBC} exists, and a reference pressure is specified, the solution ($p$) is continuously dependent on the data (i.e., $f$ and ${g}_n$).
		This means that small (bounded) perturbations to the data result in small (bounded) changes in the solution.\footnote{The error estimation in \citet{pan2016error} is built on the proof of the continuous dependence of the data of the Poisson equation.} 
		
		In the discrete setting, the sensitivity of the solution to small perturbations in the data can be quantified by the condition number of $\bm{\mathcal{L}}_N$. 
		Here, $\bm{\mathcal{L}}_N$ is rank-deficit by one, and $\text{Cond}(\bm{\mathcal{L}}_N) \to \infty$. 
		Despite this, the only zero singular value of $\bm{\mathcal{L}}_N$ is rooted in the non-uniqueness issue of \eqref{discrete neumann problem}. 
		Like the continuous case, if we apply some regularization, for example, by enforcing mean zero data and mean pressure (i.e., $\langle{\vb{b}}\rangle={0}$ and $ \langle{\vb{p}}\rangle=p_0$, guaranteeing existence and uniqueness), we can effectively remove the only zero singular value from $\bm{\mathcal{L}}_N$. 
		By doing so, the condition number of the resulting matrix is finite, and the solution to \eqref{discrete neumann problem} is continuously dependent on the data.

		\begin{table}[hbt]
			\color{black}
			\caption{\color{black} A summary of the ill-posedness issues of the Neumann problem of the Poisson equation in the continuous and discretized settings with heuristic interpretation and some remedies.}
			\label{tab: ill-possed problem}
			\centering
			\begin{tabular}{@{}p{0.1\textwidth}p{0.24\textwidth}p{0.28\textwidth}p{0.3\textwidth}@{}}
				\toprule
				& Existence                         & Uniqueness                                & Continuous dependence                                                  \\ \midrule
				\multirow{2}{*}{\makecell[l]{Continuous\\setting}}   & \multicolumn{3}{c}{(Potential) problems for the ill-posed \eqref{eq: PPE with NBC}}                                                                                                   \\ \cmidrule(l){2-4} 
				&\makecell[l]{Violating \eqref{eq: compatibility condition clean}} & \makecell[l]{Having more than \\ one solution to $p$}                  & \makecell[l]{Unbounded variation in $p$\\ when $f$ and/or $g_n$ are perturbed}                    \\ \midrule
				\multirow{2}{*}{\makecell[l]{Discretized \\setting}}  & \multicolumn{3}{c}{(Potential) problems for the ill-posed \eqref{discrete neumann problem}}                                                                                                           \\ \cmidrule(l){2-4} 
				& \makecell[l]{ $\vb{b}_N \notin\text{Col}(\mathcal{L}_N)$,\\ \eqref{discrete neumann problem} is inconsistent}             & \makecell[l]{$\mathcal{L}_N$ is not one-to-one,\\ $\vb{p}_0$ is not specified, \\ \eqref{discrete neumann problem} is underdetermined} & \makecell[l]{$\text{Cond}(\bm{\mathcal{L}}_N) \to \infty$, \\unstable computation}                                              \\ \midrule
				\makecell[l]{Heuristics}                     & Flux and generation are not balanced   & Pressure is a relative value              & Small perturbations in input causing drastic changes in output is nonphysical \\ \midrule
				\makecell[l]{Potential\\ remedies} & \makecell[l]{LS (or MNLS)\\ Sects.~\ref{sec: underdetermined regu}~\&~\ref{sec: one point DBC}\\ Remarks~\ref{rmk: well-posed problem} \& \ref{rmk: guaranteed compatibility}\\ Notes~\ref{note: div therom}, \ref{note: enforce mean-zero b} \&   \ref{note: Compute MNLS}}             &\makecell[l]{MN (or MNLS) \\Sects.~\ref{sec: underdetermined regu} \& \ref{sec: one point DBC} \\ Remark~\ref{rmk: ODI = MN or MNLS solution} \& \ref{rmk: well-posed problem} \\ Notes ~\ref{note: uniqueness by auxiliary conditions} \&~\ref{note: Compute MNLS}}    & \makecell[l]{MN or MNLS \\Regularization~(e.g., Appx.~\ref{sec: appendix: Tikonov and CG for SVD})}                                                      \\ \bottomrule
			\end{tabular}
		\end{table}

		\color{black}
		
		\section{Ill-posed Neumann Problem, Conjugate Gradients, and Minimum Norm Least Squares Solution}
		\label{sec: underdetermined regu}
		
		One may wonder how \citet{zigunov2024one} solved the general ill-posed system of \eqref{discrete neumann problem} outlined in Sect.~\ref{sec: Well-posedness of the Neumann problem}. By answering this question, we give a precise interpretation of the output from RPR-ODI. 
		
		The key to the success of \citet{zigunov2024one} lies in (i) their choice of stencil outlined in Sect.~\ref{sec: ODI = PPE}, and (ii) the use of the Conjugate Gradient (CG) algorithm to solve the resulting linear system.   
		For (i), the compatibility of the Poisson equation is guaranteed and the resulting system is consistent, thus, a solution exists.
		For (ii), running CG on a consistent, symmetric, and semi-definite system converges \textit{towards} the Minimal Norm (MN) solution. 
		By doing so, the uniqueness issue of the problem is resolved.
		Classic analysis on this topic can be found in \citet{kammerer1972convergence,hestenes2012conjugate}.

		This means that the \textit{results} from the ODI methods are equivalent to \textit{pursuing} a `solution' to the following constrained optimization problem:
		\begin{align}
			\label{eq: under determined optimization}
			\begin{split}
				\min_{\vb{p}} ~   & \mathcal{J} = \|\vb{p} \|_2^2 \\
				\text{s.t.} ~ & \bm{\mathcal{L}}_N \vb{p} = \vb{b}_N.
			\end{split}
		\end{align}
		\textcolor{black}{When the compatibility condition \eqref{eq: compatibility condition clean} is respected, which is the case for \citet{zigunov2024one}, the constraint in \eqref{eq: under determined optimization} is satisfied, and this optimization problem has an exact solution. 
			Appendix~\ref{sec: appendix: LS, MN, and MNLS solutions} provides more discussions on the MN solution.}
		
		In the case where the compatibility condition is not satisfied, for instance, due to inconsistent computation, and/or even numerical errors, 
		$\tilde{\vb{b}}_N$ can corrupt the constraints in \eqref{eq: under determined optimization} as described in Sect.~\ref{sec: Well-posedness of the Neumann problem}, and the exact solution to the above optimization problem does not exist.\footnote{\textcolor{black}{Hereafter, we will use the $\tilde{[\cdot]}$ accent to emphasize the error-contaminated quantities when necessary.}} 
		As a result, directly running CG on the inconsistent system may diverge depending on the starting point, stopping criteria, and the specific implementation of the CG methods used. 
		Despite this, one could still pursue a $\hat{\vb{p}}$ that \textcolor{black}{best satisfies the ``constraints'' $\bm{\mathcal{L}}_N \hat{\vb{p}} \approx \tilde{\vb{b}}_N$ (e.g., minimizing the residual $\vb{r} = \bm{\mathcal{L}}_N \hat{\vb{p}}  - \tilde{\vb{b}}_N $ in the sense of some norm), while minimizing $\hat{\vb{p}}$.} 
		\textcolor{black}{The most straightforward approach to such a problem is to pursue a Minimal Norm Least Squares (MNLS) solution. 
			This understanding allows us to provide a precise interpretation of the outcome of the ODI methods, which is the second major result of the current work:}
		
		\color{black}
		\begin{rmk}[ODI pursues MN or MNLS solution]
			\label{rmk: ODI = MN or MNLS solution}
			The result from ODI methods is either a Minimal Norm (MN) solution or a Minimal Norm Least Squares (MNLS) solution to a Neumann problem of PPE depending on the consistency of the corresponding linear system \eqref{discrete neumann problem}.
		\end{rmk}
		\indent The subtle differences between the MN solution to an semi-definite consistent system and the MNLS solution to an semi-definite inconsistent system, as well as the familiar Least Squares (LS) solutions are summarized for convenience in Appendix~\ref{sec: appendix: LS, MN, and MNLS solutions}. 
		Here, we want to emphasize that the numerical implementation by \citet{zigunov2023fast} implicitly encodes the compatibility condition by enforcing $\nabla p = \bm{g}$ pointwisely (see Sect.~\ref{sec: ODI = PPE}).
		Ignoring potential numerical nuances, OS-MODI---and perhaps conventional ODI methods such as RPR-ODI---converge \textit{towards} the MN solution. 
		\color{black}
		
		
		\textcolor{black}{Despite this, it is challenging to precisely comment on the extent to which conventional ODI methods recover the actual MN or MNLS solution.
			Regardless, it is interesting to see why a MNLS solution is a desirable solution from varying perspectives.}
		The MNLS solution to the inconsistent semi-definite system of \eqref{discrete neumann problem} can be expressed as 
		\begin{equation}
			\label{eq: pseudo-inverse}
			\hat{\vb{p}} =  \bm{\mathcal{L}}_N^\dagger \tilde{\vb{b}}, 
		\end{equation}
		where $\bm{\mathcal{L}}_N^\dagger$ is the Moore-Penrose pseudo-inverse of $\bm{\mathcal{L}}_N$. 
		This solution is an MNLS solution and carries several remarkable properties.
		We provide a brief interpretation, in the context of pressure field reconstruction, following the classic results of generalized inverse \citep{campbell2009generalized,gallier2020linear} with the remarks below:

		\begin{rmk}[least squares solution] 
			\label{rmk: LS solution}
			$\hat{\vb{p}}$ is a least squares solution to $\bm{\mathcal{L}}_N \vb{p} =\tilde{\vb{b}}_N$. This means that $\hat{\vb{p}}$ best matches the \textcolor{black}{potentilly incompatible} data $\tilde{\vb{b}}_N$ with respect to the $L^2$-norm.
		\end{rmk}
		
		\color{black}
		\indent $\hat{\vb{p}}$ minimizes the squared error  between $\tilde{\vb{b}}_N$ and $\bm{\mathcal{L}}_N \vb{p}$. 
		It is self-evident that this is beneficial. 
		It is also interesting to note that, in general, an (ordinary) least squares solution facilitates a Best Linear Unbiased Estimator (BLUE) provided the Gauss-Markov assumptions are satisfied. 
		However, the MNLS solution to \eqref{discrete neumann problem} fails to achieve the `optimality' of the BLUE as it breaks several critical assumptions (e.g.,  $\tilde{\vb{b}}_N$ is not generally contaminated by \textit{uncorrelated mean-zero random noise with constant variance} and $\bm{\mathcal{L}}_N$ is not full rank. See Sect.~\ref{sec: robustness of ODI/PPE} for more discussions).
		
		\color{black}

		
		
		\begin{rmk}[solution of minimal correction]
			\label{rmk: minmal correction}
			$\hat{\vb{p}}$ is an outcome of applying a minimal correction to the corrupted data $\tilde{\vb{b}}_N$, achieved by orthogonal projection. 
		\end{rmk}
		
		This can be explained by observing the fact that
		$\hat{\vb{p}}$ is also an \textit{exact} solution of $\bm{\mathcal{L}}_N \vb{p} = \bm{\mathcal{L}}_N \bm{\mathcal{L}}_N^\dagger \tilde{\vb{b}}_N$.
		$\mathcal{\vb{P}} = \bm{\mathcal{L}}_N \bm{\mathcal{L}}_N^\dagger$ projects $\tilde{\vb{b}}_N$ into the column space of $\bm{\mathcal{L}}_N$ satisfying the compatibility condition, and removing the \textcolor{black}{error---rooted from experimental measurement and/or numerical implementation---}in $\tilde{\vb{b}}_N$ that violates the compatibility condition. 
		\textcolor{black}{This projection perspective is an alternative way to look at the least squares as discussed in Remark~\ref{rmk: LS solution}.} 
		In addition, $\mathcal{\vb{P}}$ is orthogonal; thus, the erroneous component removed from $\tilde{\vb{b}}_N$ by $\mathcal{\vb{P}}$ is the smallest possible with respect to the $L^2$-norm. 
		The minimal action of correction is often a good, if not the best choice when no additional assumption is made and no \textit{prior} is available.

		\begin{rmk}[minimal norm solution] 
			\label{rmk: min norm solution}
			$\hat{\vb{p}}$ is a minimal norm solution. That is, $\hat{\vb{p}}$ has the lowest energy (measured by the $L^2$-norm) among all other possible least squares solutions. 
			This recovers the intention of \eqref{eq: under determined optimization}, and automatically requires that $\hat{\vb{p}}$ be mean-zero.
		\end{rmk}
		
		Here, the mean-zero property of the minimal norm solution is a result of $\text{Null}(\bm{\mathcal{L}}_N^\intercal)=\bm{1}$. As mentioned in Sect.~\ref{sec: uniq}, the solution $\vb{p}$ is unique up to a constant. Enforcing minimal norm  in this case requires $\bm{1}^\intercal\vb{p}=0$, and as a result, $\vb{p}$ must be mean zero.

		
		
		
		\color{black}

		\begin{rmk}[well-posed problem] 
			\label{rmk: well-posed problem}
			$\hat{\vb{p}}$ always exists and is unique.
		\end{rmk} 
		
		The challenge of the original ill-posed Neumann problem for the Poisson equation is no longer an issue, and the corresponding MNLS problem is now well-posed.
		The significance of this feature is self-evident.
		
		\begin{rmk}[no curl-free correction] 
			\label{rmk: not curl-free}
			\textcolor{black}{$\hat{\vb{p}}$ is precisely a minimum norm (least squares) solution, and it does not perform curl-free correction to $\tilde{\bm{g}}$ per se.}
		\end{rmk} 
		
		The ODI methods are motivated by the gradient theorem, which states that the line integral along the gradient field of a scalar (e.g., contaminated pressure gradients) should be path-independent.
		\textcolor{black}{
			Although the motivation of gradient theorem demands curl-free pressure gradients ($\nabla \times \tilde{\bm{g}} = \bm{0}$), the MN or MNLS solution itself does not provide any curl-free correction to $\tilde{\bm{g}}$.
			In other words, with respect to curl-free correction, ODI is neither more nor less effective than PPE, as $\text{ODI}\equiv\text{PPE}$.
			A brief elaboration on how ODI and common PPE solvers perform curl-free correction in the domain but \textit{not} on the boundary can be found in Appendix~\ref{sec: appendix: curl-free correction}. 
		}

		\section{On the Robustness of ODI/PPE}
		\label{sec: robustness of ODI/PPE}
		The above interpretation in Sect.~\ref{sec: underdetermined regu} illustrates the ODI methods' actual intention and the solution's exact sense: \textcolor{black}{an MN or MNLS} solution. 
		Despite this, one may still wonder, what is the fundamental mechanism behind the robustness of the ODI (and the equivalent PPE)?
		Perhaps the most obvious explanation to the robustness of the ODI is that the ensemble averaging practice of the traditional ODI algorithms work to cancel out the random noise.
		However, this is one way---but not necessarily a fundamental way---to look at it: the matrix ODI methods in \citet{zigunov2023fast,zigunov2024one} and some examples in the current work (see Sect.~\ref{sec: validation}) do not apply any ensemble averaging and achieve the same accuracy and robustness as the conventional RPR-ODI.
		
		Extending the comprehensions in Sect.~\ref{sec: underdetermined regu}, we may gain insights into why the ODI/PPE is robust against random noise from various perspectives,
		which may inspire improvements within easy reach.
		We keep notes herein as follows, in the order of importance in the authors' view, which could be subjective.  
		
		\begin{rmk}[guaranteed compatibility]
			\label{rmk: guaranteed compatibility}
			\textcolor{black}{OS-MODI (and RPR-ODI) implicitly computes the data to guarantee the compatibility of the Neumann problem of the Poisson equation.} 
		\end{rmk}
		\indent\textcolor{black}{Upon proper discretization of the PPE (see Sect.~\ref{sec: ODI = PPE}), the compatibility condition is satisfied and ensures the existence of the solution (see Sect.~\ref{sec: Well-posedness of the Neumann problem} and Remark~\ref{rmk: well-posed problem}).
			This is not only important for Neumann problem as we have discussed so far, but it is also critical when Dirichlet conditions are prescribed.
			Implementing any Dirichlet condition mathematically guarantees a compatible problem; however, it is still required to ensure the compatibility of the underlying Neumann problem (see Sect.~\ref{sec: one point DBC}) for a robust computation.
			This is essentially what ODI methods do, and it is one key to their success.}

	\begin{rmk}[intrinsic low pass filter]
		The intrinsic low-pass filter effect of integrating the Laplacian or pressure gradients eliminates noise in the data $\tilde{f}$ and $\tilde{\bm{g}}$ respectively. 
	\end{rmk}
	\indent
	This smoothing property is independent of the numerical method. It is rooted in the nature of integration (i.e.,  $\nabla^{-2}$ for $f$ or $\nabla^{-1}$ for $\bm{g}$, and see \citet{de2012instantaneous,faiella2021error,li2024error} for some discussions) and one may recall the heuristic that the integration of a noisy signal is almost always smoother than the original noisy signal.

	\color{black}
	\begin{rmk}[`almost' BLUE]
		\label{note: blue}
		$\hat{\vb{p}}$ from LS could be BLUE, provided the Gauss-Markov assumptions are satisfied \citep{henderson1975best}. OS-MODI and RPR-ODI are not necessarily BLUE, but somewhat close.
	\end{rmk} 
	\noindent

	Looking through the statistical lens, we can extend Remark~\ref{rmk: LS solution} and gain additional insights and inspire readily implementable improvements to the ODI/PPE by some classic results.
	Gauss-Markov assumptions requires $\bm{\mathcal{L}}_N$ being full column rank, and the errors in 
	$\tilde{\vb{b}}_N$ is uncorrelated mean-zero random noise. 
	These assumptions are fragile in our context. Since $\bm{\mathcal{L}}_N$ is rank-deficient, solving \eqref{discrete neumann problem} using MNLS yields a linear unbiased estimator. However, it is not necessarily the best, meaning that it unfortunately may not be a minimum variance estimator (i.e., BLUE), which is typically expected from a least squares solution.
	One way to resolve this `issue' is to modify and concatenate additional independent rows to  $\bm{\mathcal{L}}_N$ and $\tilde{\vb{b}}_N$, which can be from some auxiliary conditions. 
	\color{black}
	
	Despite not necessarily being the best in terms of variance control, potential \textit{unbiasedness} of the solution \eqref{eq: pseudo-inverse} is critically attractive, which implies that $\hat{\vb{p}}$ `faithfully' carries the statistics of the noise in $\tilde{\vb{b}}_N$, and
	$\mathbb{E}[{\hat{\vb{p}}}] = \vb{p}$  \textcolor{black}{if the noise in $\tilde{\vb{b}}_N$ and the true value of the pressure are both zero mean.} In this case, the covariance of the estimate is of the form 
	\begin{equation}
		\label{eq: var}
		\text{Cov}[{\hat{\vb{p}}}] = \sigma_b^2 (\bm{\mathcal{L}}_N^\dagger)^\intercal\bm{\mathcal{L}}_N^\dagger,
	\end{equation}
	where $\sigma_b^2$ is the standard deviation of the noise in $\tilde{\vb{b}}_N$. 
	Equation~\eqref{eq: var} effectively gives an uncertainty quantification, given that $\sigma_b^2$ can be assumed as a constant and is accessible. 
	
	The \textit{consistency}\footnote{Consistency here is in the context of statistics. An estimator of a given parameter is said to be consistent if it converges in probability to the true value of the parameter as the sample size tends to infinity.} of the solution suggests that larger data (e.g., from higher-resolution experiments) would allow us to better recover the unknown true value of the pressure field (an alternative justification by Fourier analysis can be found in \citet{li2024error}).
	Interestingly, the consistency of the solution of \eqref{eq: pseudo-inverse} immediately recovers the \textcolor{black}{intention and major result} of \citet{liu2020error}: if the pressure gradient is contaminated by point-wise independent mean-zero noise, the error in the pressure reconstruction converges to zero as the resolution of the data increases to infinitely high.\footnote{Note, consistency is not necessarily an error estimate or uncertainty quantification.}
	
	\color{black}
	In realistic PIV data, the error in the pressure gradients is a non-linear function of both the measurement error and the measured flow field, and both the measurement and the error can be spatially correlated and varying.
	This makes the error in $\tilde{\vb{b}}_N$ nonlinearly flow-dependent, inhomogeneous, spatially correlated, and not necessarily
	mean-zero, which break the Gaussian-Markov assumptions in many ways.
	\color{black} 
	Thus, BLUE is not necessarily guaranteed by ODI methods (including RPR-ODI and matrix-ODI methods) and the equivalent PPE-based methods.
	Moreover, the uncertainty quantification suggested in \eqref{eq: var}, strictly speaking, is not valid.
	Instead, without careful treatment and further improvement, ODI or PPE solution by \eqref{eq: pseudo-inverse} achieves a statistically sub-optimal solution.
	
	This understanding also allows us to comment on the impact of how noise is introduced when generating synthetic data for validation. 
	One can introduce noise to the pressure gradient field ($\bm{g}$) or to the velocity field ($\bm{u}$), as we will soon demonstrate in Sects.~\ref{sec: ODI vs. PPE} and \ref{sec: Compatibility of PPE}, respectively.
	When point-wise Independent and Identically Distributed (IID) random noise is added to $\bm{g}$, the solution from RPR-ODI and PPE is BLUE if boundary conditions are properly treated and $\bm{\mathcal{L}}_N$ is modified to be full column rank. 
	This practice is proper for investigating the properties of a pressure solver itself; however, it artificially accommodates the requirements for BLUE and favors BLUE-orientated solvers, despite being not realistic for validation.
	\textcolor{black}{Real noise in PIV and similar velocity measurements likely violate the Gaussian-Markov assumptions. 
		Rigorous validation and verification of the performance of the pressure solvers shouldn't use oversimplified errors, which often lead to unrealistically favorable results. 
		Proper design of artificial error for benchmarking is an intriguing topic itself, we will leave it for future studies.} 
	
	
	From this estimation perspective, we can identify some obvious strategies to improve ODI/PPE-based pressure solvers. 
	When the error in the PIV is considered flow-dependent (usually not IID), and additive to $\bm{u}$---instead of additive to $\bm{g}$---it is possible to use weighted pseudo-inverse to achieve BLUE \citep{henderson1975best}. 
	This approach is already demonstrated in \citet{zhang2020using}.
	Their tests show an approximate 20\% improvement in accuracy compared to using regular pseudo-inverse. 
	
	For an alternative improvement strategy, one may consider the classical idea of trading off the requirement of linearity and/or unbiasedness for a more accurate estimate of $\vb{p}$, which is possible by the familiar Tikhonov regularization \citep{hoerl1970ridge,hastie2009elements}.
	In fact, improvement by regularization is almost always guaranteed upon a proper choice of regularization method and parameters \citep{hastie2009elements}. 
	We will leave this topic for future discussions.

	\textcolor{black}{\textcolor{black}{From this regression perspective, we can more directly comment on why the RPR-ODI outperforms the older versions of the ODI.  Different versions of ODI methods can be reviewed as  generalized least squares with various weights.}
		Earlier versions of ODI methods implicitly assign less-than-optimal weights due to the prescribed shape and size of the virtual boundary, while RPR-ODI adopts a uniform weighting scheme under ordinary least squares. In both cases, however, the weights are governed by solver hyperparameters rather than being informed by data quality or flow characteristics, highlighting opportunities for improvement.} 

	From the above remarks in this section, we may finally see why the ODI has been reported to be robust against random noise and how to make further improvements.

	\section{Poisson problem with singular point Dirichlet condition}
	\label{sec: one point DBC}
	
	
	%
	%
	
	One common approach to regularize the Neumann problem of the Poisson equation is by simply introducing Dirichlet boundary conditions. Prescribing a Dirichlet condition at a minimum of one point on the boundary or in the domain guarantees the existence of a unique solution, and as a result, is often used in practice. The linear system corresponding to the PPE with a one-point Dirichlet condition is
	\begin{equation}
		\label{eq: one point DBC system}
		\bm{\mathcal{L}}_1 \vb{p} = \vb{b}_1,
	\end{equation}
	where $\bm{\mathcal{L}}_1$ and $\vb{b}_1$ can be constructed by replacing one line of the original underlying Neumann problem of \eqref{discrete neumann problem} with $p(\bm{x}_0) = p_0$. 
	By doing so, $\bm{\mathcal{L}}_1$ is invertible, and the solution to \eqref{eq: one point DBC system} always exists and is unique, regardless, of the error in $\vb{b}_1$.

	Despite this well-posedness, large reconstruction errors may still arise if the underlying Neumann problem does not itself satisfy the compatibility condition.
	This phenomenon can be seen in some Poisson solvers with point Dirichlet conditions where the solution floats with a spike (high slope around) at the Dirichlet point. In this case, any `generation' ($\tilde{f}$) in the domain that is not balanced by the total `flux' ($\tilde{g}_n$) through the Neumann boundaries is forced through the Dirichlet point, which is the only location where the pressure gradient is not specified. 
	This can lead to excessively high localized `flux', and a floating pressure reconstruction.
	This is likely a reason for the poor performance of some PPE-based pressure field reconstructions observed in the literature and underscores how important the proper numerical implementation of solver stencils, and the computation of $\tilde{f}$ from $\tilde{g}_n$ is for obtaining a valid reconstruction. 
	We will elaborate on this idea more in the next sections (e.g., see Note~\ref{note: div therom} and the tests in Sect.~\ref{sec: Compatibility of PPE}).

	\section{Guidelines and Notes for Practical Computation}
	\label{sec: Computational notes}
	
\begin{figure}[!bthp]
		\includegraphics[width=0.85\textwidth]{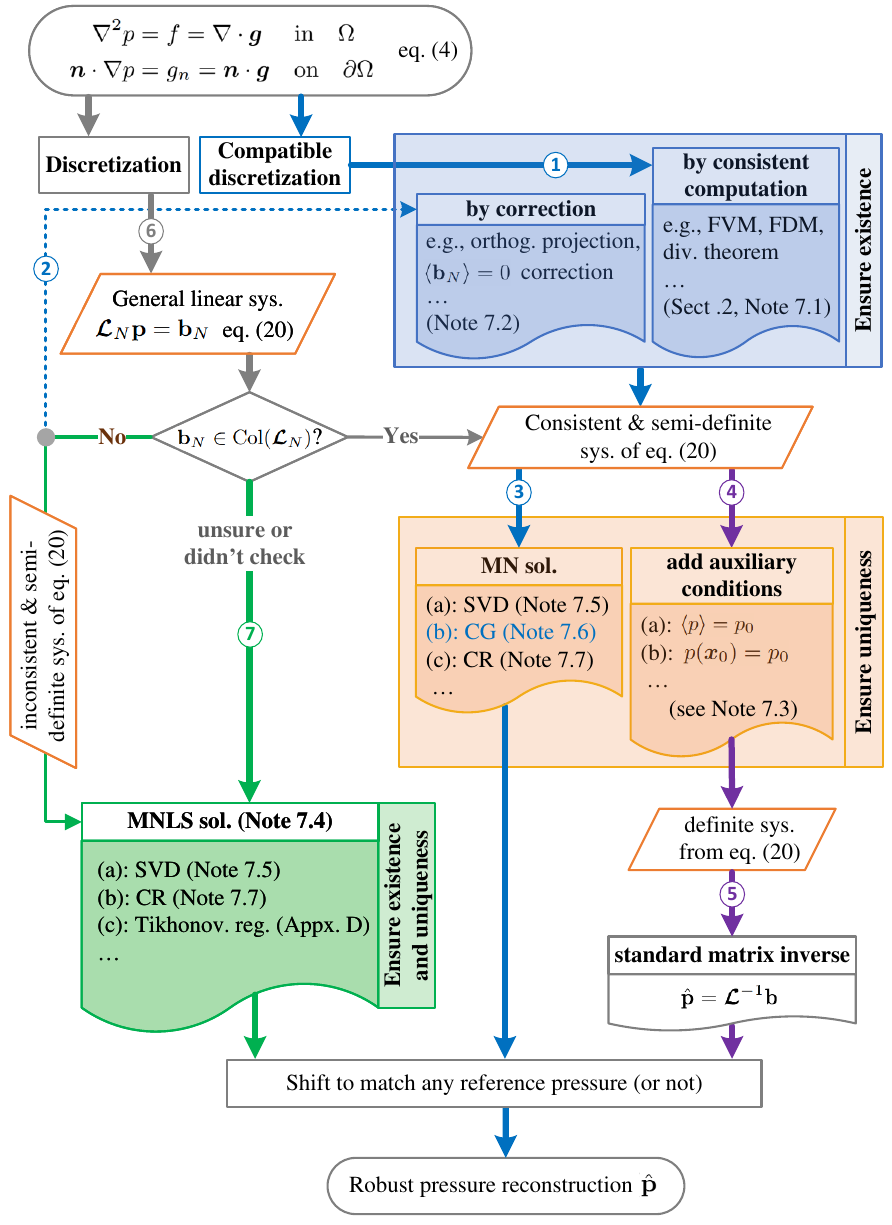}
		\centering
		\caption{Road map for robust pressure field reconstruction based on the Neumann problem of Poisson equation. Three routes are recommended: \circledtext{1}$\to$\circledtext{3}, \circledtext{1}$\to$\circledtext{4}$\to$\circledtext{5}, and \circledtext{6}$\to$\circledtext{7}. Some key pathways with abbreviations are explained as follows:    Pathway~\circledtext{1} indicates compatible discretization by consistent computation, pathway \circledtext{4} indicates ensuring uniqueness by adding auxiliary conditions to a consistent and semi-definite system of \eqref{discrete neumann problem}, and \circledtext{5} indicates solving a definite system from \eqref{discrete neumann problem} by standard matrix inverse.}
		\label{fig: flow chart}
	\end{figure}

	\color{black}
	In this section, we provide some general guidelines and notes for robust pressure reconstruction based on the Neumann problem of Poisson equations for practitioners. 
	The flow chart in Fig.~\ref{fig: flow chart} shows three routes leading to robust pressure field reconstruction, with pathways labeled by circled numbers. 
	
	
	The first two routes (\circledtext{1}$\to$\circledtext{3} or \circledtext{1}$\to$\circledtext{4}$\to$\circledtext{5}) rely on pursuing a unique solution after ensuring the existence of the solution by establishing a compatible system for the Neumann problem.
	As discussed in Sects.~\ref{sec: Well-posedness of the Neumann problem} and~\ref{sec: one point DBC}, ensuring the compatibility of the (underlying) Neumann problem of PPE is critical regardless of whether the Dirichlet conditions are prescribed or not.
	A consistent system can be achieved by a compatible discretization---e.g., using finite volume method or cell-centered finite difference as discussed in Sect.~\ref{sec: ODI = PPE}, leveraging divergence theorem (see Note~\ref{note: div therom},  pathway~\circledtext{1} in Fig.~\ref{fig: flow chart}).
	Alternatively, if an improper discretization does not necessarily lead to a compatible system, one can use some regularization so that the corrected data ($\check{\vb{b}}_N$) is in the column space of $\bm{\mathcal{L}}_N$ (see Note~\ref{note: enforce mean-zero b} and pathway \circledtext{6}$\to$\circledtext{2} in Fig.~\ref{fig: flow chart}). 
	Both the practice of \circledtext{1} and \circledtext{2} result in a consistent and semi-definite system of \eqref{discrete neumann problem}. 
	
	After this semi-definite system is constructed, there are several ways to solve for a \textit{unique} pressure field.
	One choice is to pursue the MN solution, which can be achieved using e.g., (a) Singular Value Decomposition (SVD), (b) Conjugate Gradient (CG), or (c)~Conjugate Residual (CR) indicated by pathway~\circledtext{3}. 
	It is worth noting that MATLAB's \texttt{mldivide()} can also provide fast computation for MN solution for a consistent system.  
	If using CG,  route \circledtext{1}$\to$\circledtext{3}(b) is the choice of OS-MODI in \citet{zigunov2024one}.
	Alternatively, one can introduce some auxiliary condition to make the solution to the problem unique. 
	For example, one can (a) enforce any mean value of pressure (i.e., replacing one line of \eqref{discrete neumann problem} with $\langle p \rangle =\frac{1}{m}\vb{1}^{\intercal} \vb{p} = p_0$), or (b) prescribe a one-point Dirichlet condition in the domain or on the boundary (i.e., replacing one line of \eqref{discrete neumann problem} with $p(\bm{x}_0) = p_0$).
	This practice results in a definite system, which is always consistent, and the underlying Neumann problem is compatible as it is developed through \circledtext{1} or \circledtext{2}. 
	This system can be solved by the standard matrix inverse (i.e., $\hat{\vb{p}} = \bm{\mathcal{L}}^{-1} \vb{b}$). 
	This strategy is indicated by the route \circledtext{1}$\to$\circledtext{4}$\to$\circledtext{5}.
	
	Once a solution through routes  \circledtext{1}$\to$\circledtext{3} or \circledtext{1}$\to$\circledtext{4}$\to$\circledtext{5} is achieved, one can shift the solution to match any reference pressure (or not), and they should yield identical solutions.
	We recommend these routes as they are clear, flexible, and efficient.


	The third strategy resolves all of the ill-posedness issues with the Neumann problem of PPE in one go by pursuing the MNLS solution.
	If the consistency of the Neumann problem is unknown, or somehow the potential incompatibility is difficult to check and/or correct, pursuing an MNLS solution to \eqref{discrete neumann problem} is a safe choice (see Note~\ref{note: Compute MNLS}, and  route~\circledtext{6}$\to$\circledtext{7}) despite its high computational cost for large systems.

	The computation of the pseudo-inverse or MNLS solution for an inconsistent semi-definite system is nontrivial.
	We will sample some methods including the SVD-based solution (Note~\ref{note: use svd for mnls}), as it is \textcolor{black}{exact and} foundational to the theories of pseudo-inverse.
	We will also discuss CG as an intriguing example since it is commonly used for large, symmetric, semi-definite problems and is used by \citet{zigunov2024one}; however, it is risky to use for an inconsistent system. 
	We will also recommend or discuss some other methods for practical computation of MNLS, including using CR solver (see Note~\ref{note: use CR}), Tikhonov regularization (see Note~\ref{note: use CR}), or CG with caution (see Note~\ref{note: early-stopping for CG}).
	Given the same data $\bm{g}$, through the above three routes, i.e., \circledtext{1}$\to$\circledtext{3}, \circledtext{1}$\to$\circledtext{4}$\to$\circledtext{5}, and \circledtext{6}$\to$\circledtext{7}, results in the same solution (up to some numerical tolerance).
	
	\color{black}
	
	\begin{nt}[compatibility by consistent computation]
		\label{note: div therom}
		One should ensure a consistent linear system for the Neumann problem by compatible computation.
	\end{nt}
	\indent
	There are several ways to achieve a consistent system for the Neumann problem of PPE. 
	For example, one can compute the data $\tilde{f}$ exactly as $\tilde{f} = \nabla \cdot \tilde{\bm{g}}$ and use $\bm{n}\cdot \nabla p =\tilde{g}_n = \bm{n}\cdot\tilde{\bm{g}}$ as Neumann boundary conditions with a consistent numerical scheme.  
	This practice ensures that $\tilde{g}_n = \bm{n}\cdot\tilde{\bm{g}}$ and $\tilde{f} = \nabla \cdot \tilde{\bm{g}}$ are always compatible by the divergence theorem, no matter how much $\tilde{\bm{g}}$ is contaminated.
	In the discretized context, this corresponds to using a consistent numerical scheme and stencil (e.g., finite volume method or cell-centered finite difference method as discussed in Sect.~\ref{sec: ODI = PPE}).

	This principle suggests a simple guideline for best practice.  
	If more accurate or better Neumann boundary conditions are available (e.g., $g_n=0$ at a wall or in the far field), one should correct $\tilde{\bm{g}}$ before applying the boundary conditions, then compute $\tilde{f}$ based on the corrected $\tilde{\bm{g}}$. 
	This subtle practice could effectively improve the reconstruction performance by naturally satisfying the compatibility condition of the underlying Neumann problem of the Poisson equation.
	In the following section, we will suggest additional simple regularization strategies to ensure compatibility. 
	
	
	
	\begin{nt}[compatibility by enforcing mean-zero data]
		\label{note: enforce mean-zero b}
		Correct ${\tilde{\vb{b}}}_N$ (e.g., by enforcing mean-zero) so that the corrected data ${\check{\vb{b}}_N} \in \mathrm{Col}(\bm{\mathcal{L}}_N)$.
	\end{nt}
	\indent
	There are many possible approaches to enforce the compatibility condition,
	but the simplest choice for a discrete Neumann problem of PPE is perhaps to enforce a mean-zero $\check{\vb{b}}_{N}$ by centering the data:
	\begin{equation}
		\label{eq: b=b - mean (b)}
		\check{\vb{b}}_{N} = \tilde{\vb{b}}_N - \langle{\tilde{\vb{b}}}_N \rangle\vb{1},
	\end{equation}
	and solve the consistent system $\bm{\mathcal{L}}_N \vb{p} = \check{\vb{b}}_N$ using, for example, CG. 
	$\langle{\tilde{\vb{b}}}_N\rangle$ is the mean of the inconsistent data $\tilde{\vb{b}}_N$, and the corrected data $\check{\vb{b}}_{N}$ is mean-zero (i.e., $\langle\check{\vb{b}}_{N}\rangle = 0$). 
	This simple regularization is effective as ${\check{\vb{b}}_N} \in \text{Col}(\bm{\mathcal{L}}_N)$ and \eqref{eq: b=b - mean (b)} facilitate a consistent system (see \eqref{eq: mean b = 0} and the corresponding short analysis).
	In fact, this `trick' is not new but exists in the numerical community's folklore, which is sometimes used when dealing with high round-off errors.
	We pen it here not only for housekeeping purposes but as a reminder that this `trick' is useful in a few ways---perhaps more useful in our context, where the experimental error is much more pronounced than the round-off error, and consistent computation could be challenging to achieve. 
	\textcolor{black}{First, one can verify whether the system is consistent by checking if $\tilde{\vb{b}}_N$ is mean zero---if $\bm{\mathcal{L}}_N$ is from a proper discretization.
		Second, one can use \eqref{eq: b=b - mean (b)} to correct the inconsistency of the original linear system to ensure the existence of the solution.
	}
	
	\color{black}
	Last, we want to point out that under the proper discretization, the correction of \eqref{eq: b=b - mean (b)} facilitates an easy and fast computation of MNLS solution due to the special properties of $\bm{\mathcal{L}}_N$.
	Enforcing $\langle\tilde{\vb{b}}_{N}\rangle = \vb{1}^\intercal\tilde{\vb{b}}_{N} = 0$ is equivalent to projecting $\tilde{\vb{b}}_N$ onto $\text{Col}(\bm{\mathcal{L}}_N)$, ensuring the existence of the solution.
	This is evident by seeing that the projection of $\tilde{\vb{b}}_N$ onto $\text{Null}(\bm{\mathcal{L}}_N) = \vb{1}$ can be achieved as 
	$\frac{\vb{1}^\intercal \tilde{\vb{b}}_N }{\vb{1}^\intercal \vb{1} } \vb{1} = \frac{1}{m}\vb{1}^\intercal \tilde{\vb{b}}_N \vb{1} = \langle \tilde{\vb{b}}_N \rangle \vb{1}$, and \eqref{eq: b=b - mean (b)} removes the only components in $\text{Null}(\bm{\mathcal{L}}_N)$ from $\tilde{\vb{b}}_N$.
	The least squares solution is also an outcome of the orthogonal projection of $\tilde{\vb{b}}_N$ onto $\text{Col}(\bm{\mathcal{L}}_N)$.
	It is easy to verify that this projection is $\vb{P}=\bm{\mathcal{L}}_N\bm{\mathcal{L}}_N^\dagger$ for least squares (see also Remark~\ref{rmk: minmal correction}).
	After this data correction, the resulting linear system is asymmetric, semi-definite, and consistent.

	\textcolor{black}{Similarly, in case some discretization may lead to a system $\bm{\mathcal{L}}_N\vb{p}=\vb{b}_N$, where $\text{Null}(\bm{\mathcal{L}}_N) =\vb{w}  \neq \vb{1}$, the counterpart for \eqref{eq: b=b - mean (b)} is $\check{{\vb{b}}}_N =  \bm{\mathcal{L}}_N \bm{\mathcal{L}}_N^\dag\tilde{\vb{b}}_N =  \vb{b}_N - \frac{1}{m}\vb{w}^\intercal\tilde{\vb{b}}_N \vb{1},$ where $\frac{1}{m}\vb{w}^\intercal\tilde{\vb{b}}_N$ can be viewed as a weighted average of the elements in $\tilde{\vb{b}}_N$.}

	\color{black}
	
	Although simple and effective in enforcing compatibility, this regularization method may be somewhat `blunt' and does not guarantee the best results.
	It performs `indiscriminate' correction to $\tilde{f}$ and $\tilde{g}_n$ embedded in $\tilde{\vb{b}}_N$, and as a result, this adjustment may `correct' something that should not be corrected.
	We will demonstrate this in the following section.

	\color{black}
	
	\begin{nt}[uniqueness by auxiliary conditions]
		\label{note: uniqueness by auxiliary conditions}
		Once the existence of the solution is guaranteed, the uniqueness of the solution can be achieved by introducing extra conditions.
	\end{nt}
	To ensure a unique solution, it is necessary to anchor the pressure field, which is an inherently relative quantity. 
	Several common approaches exist, and we list a few here.
	One straightforward method is to impose Dirichlet conditions. For example, one can specify the pressure at a particular point in the domain or on the boundary (i.e., $p(\bm{x}_0) = {p}_0$),  prescribe the mean pressure across the entire field (i.e., $\langle{\vb{p}}\rangle = \frac{1}{m}\vb{1}^\intercal \vb{p} = {p}_0$), or enforce the mean pressure over a subset of the domain—for instance, (e.g., \citet{wang2023greens} imposed a zero-mean pressure on the boundary). 
	Another natural approach is to formulate the problem as an optimization task and seek a minimal norm solution, as discussed in this work. Notably, the choice of norm need not be limited to the $L^2$-norm.
	While some of these anchoring strategies may be connected or even equivalent in certain contexts, the most appropriate choice could be problem-specific. 
	A deeper investigation into this subject is left for future work.

	\color{black}

	\textcolor{black}{
		\begin{nt}[bypass ill-posedness by MNLS]
			\label{note: Compute MNLS}
			When the compatibility and uniqueness of the problem is unknown, a safe choice is to pursue the minimal norm least squares solution.
		\end{nt}
		The MNLS solution to \eqref{discrete neumann problem} always exists and is unique, regardless of the consistency of the linear system.
		There are many ways to compute MNLS. 
		For example, MATLAB's \texttt{pinv()} use SVD and  \texttt{lsqminnorm()} uses Complete Orthogonal Decomposition (COD). 
		Solving \eqref{discrete neumann problem} using the Tikonov regularization with small regularization parameter also approximately recovers MNLS solution (see Appendix~\ref{sec: appendix: Tikonov and CG for SVD} and \citet{hanke2017conjugate} for more discussions). 
		Computing MNLS for an inconsistent semi-definite system is equivalent to enforcing mean-zero data and pursuing MN solution to the resulting consistent system, as shown in Note~\ref{note: enforce mean-zero b} if~\eqref{discrete neumann problem} is developed by a proper discretization.
	}



	

	\begin{nt}[on the use of SVD for MNLS]
		\label{note: use svd for mnls}
		Singular value decomposition is foundational to analyzing the Moore-Penrose pseudoinverse, however, it can be expensive for large systems and there are other choices.
	\end{nt}
	
	The most general and robust way to compute pseudo-inverse (MN, MNLS, and LS) is perhaps to use SVD, which is also a foundational tool to study generalized inverse \citep{campbell2009generalized}. 
	Let the SVD for $\bm{\mathcal{L}}_N$ is $\bm{\mathcal{L}}_N = \vb{U}\vb{S}\vb{V}^\intercal$, then the pseudo-inverse for   $\bm{\mathcal{L}}_N$ is $\bm{\mathcal{L}}_N^\dagger = \vb{V}\vb{S}^{\dagger}\vb{U}^\intercal$, which works regardless of the shape and properties of $\bm{\mathcal{L}}_N$ (extended discussions can be found in Appendix~\ref{sec: appendix: LS, MN, and MNLS solutions}).
	In addition to being exact, SVD allows a low-rank approximation to $\bm{\mathcal{L}}_N^\dagger$, and resulting pressure reconstruction based on which can be even smoother upon a proper choice of rank.
	For practical computation of MNLS or MN for large systems, some preliminary recommendations are listed in Notes~\ref{note: early-stopping for CG}, \ref{note: use CR}, and \ref{note: use preconditioning}.
	
	


		\begin{nt}[on the use of CG for MN and MNLS solutions]
			\label{note: early-stopping for CG}
			CG converges to the minimal norm solution on a consistent and semi-definite system. However CG may converge towards the minimal norm least squares solution on an inconsistent and semi-definite system before it blows up.
		\end{nt}
		\indent
		On a consistent and semi-definite system, CG coverages \textit{to} the minimal norm solution \citep{kammerer1972convergence,hestenes2012conjugate}. 
		However, for an inconsistent and semi-definite problem of \eqref{discrete neumann problem}, using CG is risky. 
		It may converge \textit{towards} the MNLS solution and it is almost guaranteed to blow up after certain iterations \citep{lim2024conjugate}. 
		In this case, starting the CG iterations from  $\hat{\vb{p}} = \vb{0}$ and terminating the iterations early is essential. 
		
		
		CG steps starting from $\hat{\vb{p}} = \vb{0}$ angelology the SVD and can be viewed as an iterative (approximate) solver pursuing the MNLS solutions (see Appendix~\ref{sec: appendix: Tikonov and CG for SVD} and \cite{hanke2017conjugate}).
		This is presumably the practice of \citet{zigunov2024one}, which is effective in their tests; however, an optimal stopping criterion is not obvious: one should drive the residual as low as possible before the iteration diverges (see \citet{lim2024conjugate} for an in-depth discussion on the performance of CG and other common iterative solvers on an semi-definite inconsistent system.
		This recent work indicates that CG is numerically unstable for inconsistent systems and it is not guaranteed to give the actual pseudo-inverse solution).

		\begin{nt}[use a CR or MINRES solvers]
			\label{note: use CR}
			Use a proper iterative solver, such as the Conjugate Residual (CR) or a minimum residual (MINRES) solver to solve an inconsistent problem of \eqref{discrete neumann problem} to ensure a converged solution.
		\end{nt}
		
		The key advantage of CR is that the norm of the residuals is monotonically decreasing, and it converges to the normal solution for an inconsistent system. 
		Simple projection at the termination allows CR to recover the pseudo-inverse solution without diverging (see Algorithm 4 in \citet{lim2024conjugate}). 
		Some implementations of MINRES benefit from the same monotonic convergence.\footnote{The MINRES solver in \citet{lim2024conjugate} converges on an inconsistent semi-definite system, but MATLAB's \texttt{minres()} does not.} 

		\begin{nt}[use good initial guess and preconditioning]
			\label{note: use preconditioning}
			If an iterative solver is used, there are two common strategies to accelerate the computation: starting the iteration from a good initial guess and/or applying preconditioning. 
		\end{nt} 
		
		The former is particularly relevant to the reconstruction based on time-resolved data. 
		Using the pressure field of the previous instant as the initial guess for the computation of the current instant can significantly speed up the reconstruction, as the pressure fields at adjacent moments are temporarily correlated and can share a high degree of similarity. 
		An example practice can be found in \citet{Chen2024fastpressure}. 
		\textcolor{black}{However, it could be risky to start from a non-zero or a non-zero mean initial guess when using some iterative solvers when $\bm{\mathcal{L}}_N$ is from a Neumann problem. 
			For example, if the iteration is started from $\hat{\vb{p}} = \vb{p}_0$ and $\vb{p}_0$ happens to have a component in the null space of $\bm{\mathcal{L}}_N$, this component in $\text{Null}(\bm{\mathcal{L}}_N)$ may never be removed and the MN solution would not be achieved.}
		
		The latter, preconditioning, is generically effective and sometimes can achieve improved numerical stability in addition to the benefits of acceleration. 
		For example, in the following section, we show the performance of CG with incomplete LU (ILU) preconditioning as an example.

		\section{Validation and Demonstration}
		\label{sec: validation}
		
		While some of the arguments (remarks and notes) can be validated by cross-referencing the tests in the existing literature \citep{liu2006instantaneous, zigunov2023fast, zigunov2024one, liu2020error, zhang2022uncertainty,li2024error}, we provide independent and unified numerical experiments to demonstrate key points of our arguments.

		\subsection{ODI $\equiv$ PPE}
		\label{sec: ODI vs. PPE}
		In this section, we provide direct numerical evidence on the equivalency between RPR-ODI and the PPE for pressure reconstruction. 
		To do this, we use the analytical solution of the Taylor vortex \citep{charonko2010assessment,panton2006incompressible} as the ground truth data. 
		The pressure of the vortical flow, $p$, is calculated as
		\begin{equation}
			\label{lamb vortex}
			p = -\frac{\rho H^2}{64 \pi^2 \nu t^3} \exp \left(-\frac{r^2}{2 \nu t}\right),
		\end{equation}
		where $H$ represents the angular moment of the vortex, $\nu$ the kinematic viscosity of the fluid, $\rho$ the density of the fluid, $t$ the time, and $r$ the distance from the center of the vortex in polar coordinates. 
		We choose the parameters so that the characteristic length scale of the vortex is $L_0 = \sqrt{2 \nu t} = 1$ and the leading coefficient (i.e., the amplitude of the characteristic pressure for this flow) is $P_0 = \frac{\rho H^2}{64 \pi^2 \nu t^3} = 1$.
		
		The vortex is centered at $(x/L_0,y/L_0) = (-0.5,0)$ in a $2L_0 \times 2L_0$ domain. 
		The domain is discretized on a uniform $41 \times 41$ grid with a grid spacing $h/L_0 = 0.05$. 
		To generate the synthetic data for this experiment, the ground truth pressure ($p$) and the pressure gradients ($\bm{g}$) are computed based on  \eqref{lamb vortex}. 
		The synthetic data of the contaminated pressure gradients ($\tilde{\bm{g}}$) are generated by adding mean-zero Gaussian noise with a standard deviation of $\sigma_{\bm{g}} = 0.5P_0/L_0$ to the components of ${\bm{g}}$ independently.  
		Note that the characteristic pressure gradient for this flow is $P_0/L_0 = 1$ meaning this artificial noise is rather high. 
		The noisy data for the PPE is calculated using the noisy pressure gradient and the pressure is then reconstructed using both the RPR-ODI and PPE.

		For the RPR-ODI, the ray rotation increment is set to be $\Delta\alpha=0.15$~or~0.5,  and the ray spacing is $\delta r/L_0 = 0.025$, resulting in normalized ray spacing $\Delta d^* = \delta r/ h = 0.5$.
		This choice of the hyperparameters presumably facilitates a high-resolution setting ($\Delta \alpha \approx 0.1 - 0.2^\circ$ and $0.35 <\Delta d^* < 0.71$) recommended by \citet{liu2020error} for accurate reconstruction based on RPR-ODI.
		The tolerance of the RPR-ODI iteration is set to be $10^{-10}$ and the maximum iteration is $n_{\text{itr}} = 20$.
		Starting from an initial guess of $p=0$, RPR-ODI typically converges in about 10 steps.
		The computed pressure is then shifted to match the ground truth at a corner of the domain (i.e., $(x/L_0,y/L_0) = (-1,-1)$). 
		
		For the PPE, a finite difference approximation is used with all Neumann boundary conditions and the particular stencils shown in Sect.~\ref{sec: ODI = PPE}. 
		The resulting singular but consistent system of equations of \eqref{discrete neumann problem} was solved using both MATLAB's pseudo-inverse (i.e., \texttt{pinv()}) and a CG solver with maximum iteration of $n_{\text{itr}} = 41^2 = 1681$ which is the dimension of $\bm{\mathcal{L}}_N$ and tolerance being $\text{tol} = 10^{-8}$).
		\textcolor{black}{We dub these two PPE solvers as PPE-PINV and PPE-CG herein, which correspond to the routes \circledtext{1}$\to$\circledtext{3}(a) and \circledtext{1}$\to$\circledtext{3}(b), respectively, in Fig.~\ref{fig: flow chart}.} 
		Similar to the use of RPR-ODI, the solutions from PPE were shifted to match the pressure ground truth at $(x/L_0,y/L_0) = (-1,-1)$.
		The aforementioned procedure was then repeated a total of 500 times each with independently generated random noise and compared for a statistical test.
		\textcolor{black}{We want to emphasize again that the PPE-CG practice tested here is identical to running OS-MODI introduced by \citet{zigunov2024one}.}

\begin{figure}[thbp!]
			\begin{adjustbox}{addcode={\begin{minipage}{1.1\width}}{\caption{Comparison of ODI and PPE: pressure field ground truth (a), pressure reconstructed using RPR-ODI (b), PPE by pseudo-inverse (c) and CG (d), respectively.
								(g-i) error in the pressure reconstruction by comparing (b-d) with (a) respectively.
								(e) Box plots of pressure reconstruction ($\hat{p}$) and error in the pressure ($\epsilon_p$) correlation between RPR-ODI and PPE solved using pseudo-inverse respectively; (f) Box plots for the space-averaged $L^2$-norm of the error in the pressure reconstruction by RPR-ODI (A), PPE-PINV (B) and PPE-CG (C), respectively. Box plots (e~\&~f) are based on 500 independent tests. Horizontal bars in the middle of the boxes show the median while the upper and lower edges of the box indicate the 25 and 75 percentiles. The upper and lower whiskers bound the 95\% confidence intervals of the error while the diamond symbols within the boxes mark where the corresponding tests shown in (b--d and g–-i) lie within the data.       
								\label{fig: lamb vortex}}
					\end{minipage}},rotate=90,center}   \includegraphics[width=1.1\textwidth]{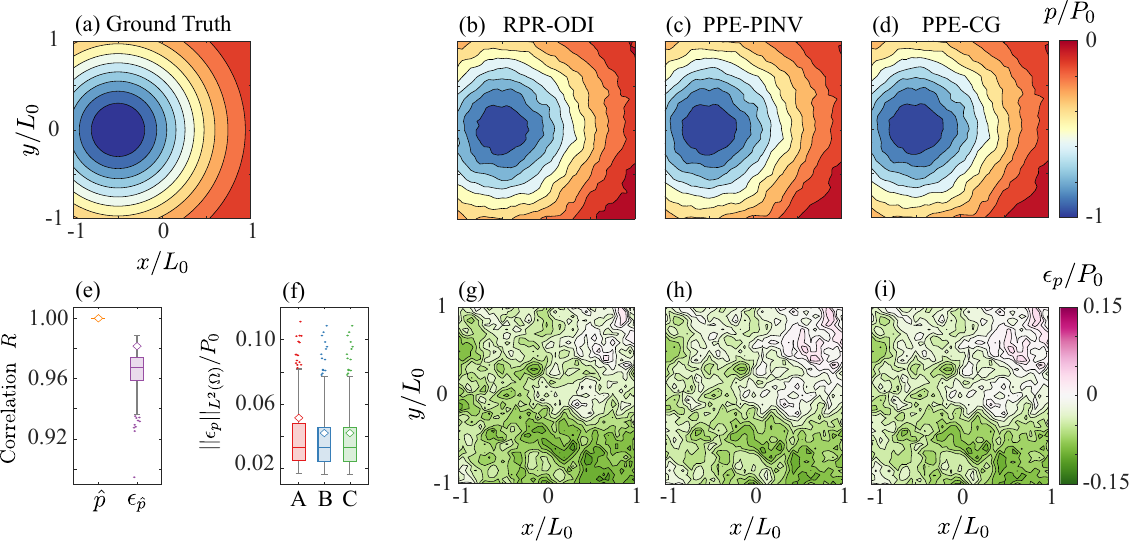}%
			\end{adjustbox}
		\end{figure}

		Figure~\ref{fig: lamb vortex} demonstrates the typical and statistical performance of the solvers. 
		In Fig.~\ref{fig: lamb vortex}(e), we show the correlations of the results (i.e., pressure reconstruction $\hat{p}$ and the error in the pressure reconstruction $\epsilon_p$) by RPR-ODI and PPE solved by Moore-Penrose pseudo-inverse. The correlations are computed as
		\begin{equation*}
			\label{eq: R}
			R = \frac{\text{Cov}[{X}_{\text{ODI}},{X}_{\text{PPE}}]}{\sqrt{\text{Var}[X_{\text{ODI}}]\text{Var}[X_{\text{PPE}}]}}, 
		\end{equation*}
		where $X$ represents the reconstructed pressure or the error in the reconstructed pressure.
		In our tests, the mean of the 500 correlations for the pressure reconstructions is $R(\hat{p}) = 0.9998$ for $\Delta \alpha = 0.5~\text{ and }~0.15$.
		This suggests that the pressure reconstructed by RPR-ODI and \color{black} PPE \color{black} solved by pseudo-inverse is practically identical, which validates the equivalency of ODI and PPE argued in Sect.~\ref{sec: ODI = PPE}.
		The mean of the reconstruction error correlations are $R(\epsilon_{p}) = 0.9653$~and~0.9654, which are also high. 
		This further supports the ODI~$\equiv$~PPE argument.
		A set of typical pressure reconstructions using RPR-ODI, PPE solved by pseudo-inverse and CG based on the same data out of 500 tests are illustrated in Fig.~\ref{fig: lamb vortex}(b--d), respectively. 
		The corresponding error maps in the reconstructed pressure field are shown in Fig.~\ref{fig: lamb vortex}(g--i), respectively. 
		Both the pressure fields and the error maps are virtually identical as expected.
		
		We also notice that $R(\hat{p})$ and $R(\epsilon_{p})$ are not precisely unity (indicating perfect similarity). 
		This may be due to the numerical implementation details of the methods. 
		For example, to conduct an RPR-ODI, one must choose extra hyperparameters such as the ray rotation angle increment, ray spacing, tolerance, and maximum number of iterations, while PPE solved by pseudo-inverse could be implemented using different stencils and different algorithms for pseudo-inverse (e.g., the results from SVD- or COD-based algorithms, or some iterative solvers such as CG or GR are not exactly the same).
		In addition, the low spatial resolution of data in these tests, adopted to accommodate the high computational cost of mass trials for RPR-ODI, may also contribute to the slight spread in the correlations.
		
		The three boxes in Fig.~\ref{fig: lamb vortex}(f) show the statistics of the reconstruction error based on 500 independent tests using RPR-ODI (A), PPE solved by SVD-based pseudo-inverse (B), and PPE solved by CG (C).
		The statistics of the errors are very similar; however, the reconstruction based on the PPE is slightly better than that from the high-resolution RPR-ODI.
		This can be seen by observing the sample mean and standard deviation of the errors in the reconstructed pressure as shown in Table~\ref{tab: mean and std}. As the resolution of RPR-ODI increases from $\Delta \alpha = 0.5^\circ$ to $\Delta \alpha = 0.15^\circ$, the sample error statistics from RPR-ODI approach the error statistics from PPE. 
		This suggests that solving the PPE is the same as taking the resolution of RPR-ODI to its limit. 
		This is also supported by \cite{zigunov2023fast} in the development of I-MODI where the number of ray integrals are taken to infinity, (i.e. $\Delta d^* \to 0$ and $\Delta \alpha \to 0$).


		\begin{table}[thbp!]
			\caption{The statistics (sample mean and standard deviation) of the reconstruction error for RPR-ODI, PPE-PINV, and PPE-CG. The hyperparameters for RPR-ODI are $\Delta d^* = 0.5$ and $\Delta \alpha = 0.5~ \text{or}~0.15$. The error distribution for the case of $\Delta \alpha = 0.5$ is shown in Fig.~\ref{fig: lamb vortex}(f).}
			\centering
			\begin{tabular}{@{}ccccc@{}}
				\toprule
				Statistics                                                      & $\Delta \alpha$ & RPR-ODI (A)           & PPE-PINV (B)                           & PPE-CG (C)                             \\ \midrule
				\multirow{2}{*}{$\mathbb{E}[||\epsilon_p||_{L^2(\Omega)}]/P_0$} & $0.5^\circ$     & $3.84 \times 10^{-2}$ & \multirow{2}{*}{$3.73 \times 10^{-2}$} & \multirow{2}{*}{$3.73 \times 10^{-2}$} \\
				& $0.15^\circ$    & $3.78 \times 10^{-2}$                   &                                        &                                        \\ \midrule
				\multirow{2}{*}{$\sigma_{\epsilon_p}/P_0$}                      & $0.5^\circ$     & $1.76 \times 10^{-2}$ & \multirow{2}{*}{$1.67 \times 10^{-2}$} & \multirow{2}{*}{$1.67 \times 10^{-2}$} \\
				& $0.15^\circ$    & $1.72 \times 10^{-2}$                     &                                        &                                        \\ \bottomrule
			\end{tabular}
			\label{tab: mean and std}
		\end{table}

		\subsection{Compatibility and Stable Computing for PPE}
		\label{sec: Compatibility of PPE}
		
		We use a snapshot of a two-dimensional Taylor-Green vortex as another `minimal example' to demonstrate the importance of having compatible data and to showcase some of the associated critical nuances for stable computation. 
		
		The two components of the velocity field of the Taylor-Green vortex are $u = U_0\sin(\pi x/L_0)\cos(\pi y/L_0)$ and $v = -U_0\cos(\pi x/L_0)\sin(\pi y/L_0),$ where $U_0$ and $L_0$ are the characteristic velocity and length, respectively.
		The pressure field is $p =  \frac{1}{2} P_0 \left[\cos(2\pi x/L_0) + \cos(2\pi y/L_0)\right],$ where $P_0 = \frac{1}{2}\rho U_0^2$ is the characteristic pressure and $\rho$ is the density of the fluid.
		In our tests, we set $L_0$, $U_0$, and $\rho$ to unity.
		The vortex is centered within a $(x/L_0, y/L_0) \in [0,1]\times[0,1]$ domain and discretized on a uniform $126\times126$ grid with a grid spacing $h/L_0 = 0.008$. 
		
		The uncontaminated pressure gradient $\bm{g}$ is calculated from the ground truth~$p$, and then the uncontaminated data $f$ is calculated from $\bm{g}$. 
		To generate synthetic corrupted data, this time, however, independent mean-zero Gaussian noise with a constant standard deviation (i.e., $\sigma_{u,v}/U_0 = 0.03$) are added to the ground truth velocity components.
		The contaminated pressure gradient $\tilde{\bm{g}}$ is calculated from the contaminated velocity $\tilde{u}$ and $\tilde{v}$.
		Lastly, the contaminated data $\tilde{f}$ is computed from the noisy pressure gradient $\tilde{\bm{g}}$. 
		

		With the synthetic data prepared, we reconstructed the pressure field by solving the PPE using one point Dirichlet condition at the corner $(x/L_0,y/L_0) = (0,0)$.
		This `canonical' simple practice results in a mathematically well-posed problem, but it carries the risk of unstable reconstruction if not handled carefully, especially when the data are contaminated.
		We organized four different tests by slightly varying the use of data and regularization to reveal the subtle importance of the compatibility and demonstrate some arguments in Sects.~\ref{sec: one point DBC} and \ref{sec: Computational notes}.
		
		For the first test (test I), we solve the PPE using  both the contaminated data ($\tilde{f}$) and the pressure gradients ($\tilde{g}_n$) on the boundaries as Neumann conditions, this will act as a baseline. 
		$\tilde{f}$ is directly computed based on $\tilde{f} = \nabla \cdot \tilde{g}$, and thus, the resulting system is compatible (see Note~\ref{note: div therom}).
		The linear system for this test and the following three, take the form of \eqref{eq: one point DBC system} and are solved using MATLAB's \texttt{mldivide()}. 
		In this case, \eqref{eq: one point DBC system} is definite and invertible, and we don't need to solve it by pseudo-inverse.
		\textcolor{black}{This is practice is the route \circledtext{1}$\to$\circledtext{4}(b) in Fig.~\ref{fig: flow chart}.}
		
		For the second test (test II), we solve the PPE using the contaminated data $\tilde{f}$, but this time using $g_n$, the uncontaminated pressure gradients on the boundary for Neumann conditions. 
		Conventional wisdom may suggest that error-free boundary conditions should yield better results. 
		However, as we will soon see, compatibility issues between $g_n$ and $\tilde{f}$ prevent this from being the case.

		For the third test (test III), we use the same contaminated data ($\tilde{f}$) and error-free Neumann boundaries (${g}_n$), but this time we correct the data ($\tilde{\vb{b}}$) such that it is mean zero before adding in the Dirichlet condition at the corner. 
		This is the regularization suggested by \eqref{eq: b=b - mean (b)} in Note~\ref{note: enforce mean-zero b}. 
		
		For the last test (test IV), we again use the contaminated data ($\tilde{f}$) and error-free Neumann boundaries (${g}_n$), however, this time we take full advantage of the high-quality gradient data on the boundary while making sure to preserve compatibility.
		To do this, we construct a corrected pressure gradient field $\tilde{\bm{g}}_c$, such that $\tilde{\bm{g}}_c$ is equal to the error-free pressure gradient $g_n$ on the boundaries, and equal to the noisy gradient $\tilde{\bm{g}}$ that yields compatible $\tilde{f}$ in the domain.
		A such $\tilde{\bm{g}}_c$ can be achieved by simply replacing $\tilde{\bm{g}}$ on the boundary with $g_n$.
		From this corrected pressure gradient $\tilde{\bm{g}}_c$, we compute the corrected data $\tilde{f}_c= \nabla \cdot \tilde{\bm{g}}_c$ per the suggestion in Note~\ref{note: div therom}.
		We then solve the PPE using the corrected data $\tilde{f}_c$ and the error-free Neumann condition $g_n$ with the one-point Dirichlet condition.
		In most cases, accurate Neumann conditions are not easy to obtain except for some special applications (e.g., internal flows enclosed by walls, where $g_n=0$ at walls could be considered accurate boundary conditions).
		However, this test serves as a hypothetical demonstration that careful setup of the computation is needed and specific best practices are often problem-dependent.

		\begin{figure}[thbp!]
			\begin{adjustbox}{addcode={\begin{minipage}{1.0\width}}
						{\caption{Comparison of incompatible vs compatible data: (a) pressure field ground truth (b--e) reconstructed pressure using $\tilde{g}_n$ and $\tilde{f}$ following Note~\ref{note: div therom}, $g_n$ and $\tilde{f}$, $g_n$ and $\tilde{f}$ with mean-zero data correction (see Note~\ref{note: enforce mean-zero b}), $g_n$ and $\tilde{f_c}$ following Note~\ref{note: div therom} respectively; (g--j) error in the reconstructed pressure field by comparing (b--e) with (a), respectively; (f) Box plot of error in the pressure reconstruction from 500 independent tests with the red (I), blue (II) green (III), and orange (IV) boxes for the statistics of the error due to the use of different methods and data (corresponding to the description for (b--e), respectively). Horizontal bars in the middle of the boxes show the median while the upper and lower edges of the box indicate the 25 and 75 percentiles. The upper and lower whiskers bound the 95\% confidence intervals of the error while the diamond symbols within the boxes mark where the corresponding error shown in (g–-j) lie within the data.
								\label{fig: taylor vortex}}
					\end{minipage}},rotate=90,center}      
				\includegraphics[width=1.4\textwidth]{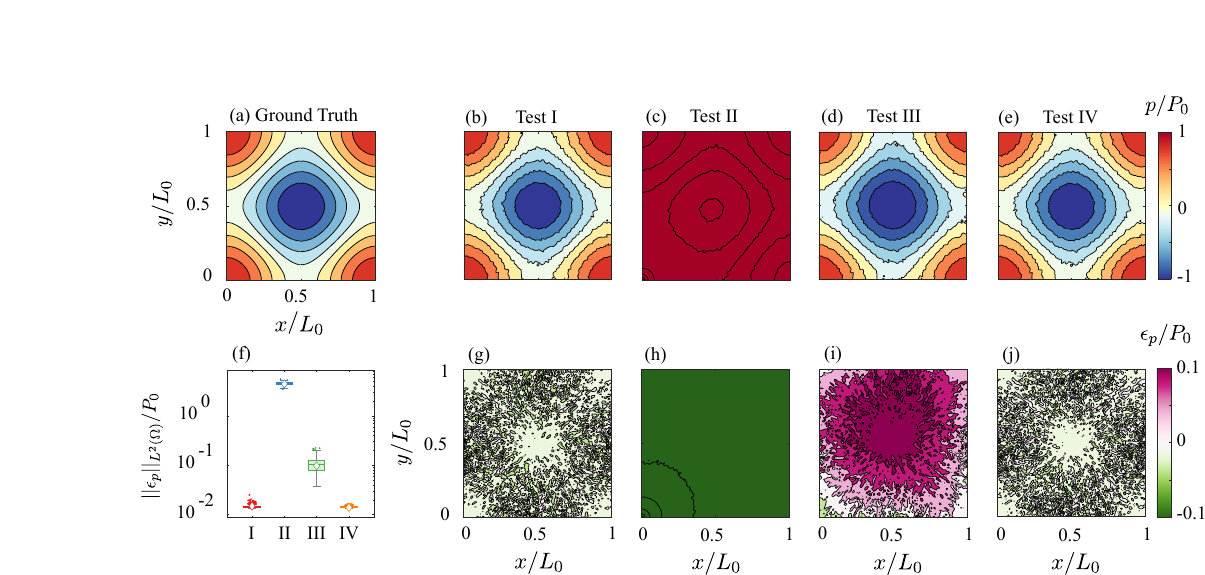}%
			\end{adjustbox}
		\end{figure}

		Figure~\ref{fig: taylor vortex} shows the results from the aforementioned four tests. 
		Each test was repeated 500 times with independently generated random noises with the same statistics introduced to the velocity field. 
		The error in the reconstructed pressure field from the four tests are shown by the four boxes in Fig.~\ref{fig: taylor vortex}(f).
		In the red box (I), the reconstructed error is rather low, despite that both the data and the boundary conditions are contaminated (see Fig.~\ref{fig: taylor vortex}(b\&g) for a typical reconstructed pressure field and the corresponding error map).
		This preferred performance is attributed to leveraging the divergence theorem as suggested in Note~\ref{note: div therom} to ensure the compatibility condition of the underlying Neumann problem.
		
		As shown in the blue box (II) in Fig.~\ref{fig: taylor vortex}(f), the errors from test~II are particularly high despite error-free boundary conditions being used. 
		This may be counter-intuitive as using more accurate data is seemingly counterproductive; however, the real reason for this `failure' is that $\tilde{f}$ and $g_n$ are not compatible: 
		$    \int_{\Omega} \tilde{f} dV - \int_{\partial \Omega} {g}_n  dS = \epsilon_c \neq 0 $.
		The unbalanced generation (the none-zero $\epsilon_c$) must go through the Dirichlet boundary (one point in our case).
		Depending on the value of $\epsilon_c$ and the length of the Dirichlet boundaries, the `flux' or the pressure gradient near the Dirichlet boundaries can be excessively high.
		In the case of our test, the pressure value at the $(x/L_0, y/L_0) = (0,0)$ corner is anchored and the steep gradient near this corner shifts the entire pressure field, leading to significant errors in the reconstruction. 
		This effect is highlighted in Fig.~\ref{fig: taylor vortex}(c\&h), where the reconstructed pressure appears to have a high `constant' bias except at the Dirichlet point, where the pressure value remains exact.

		The regularization used in test III resolves this compatibility issue and the corresponding high reconstruction error. 
		As shown in Fig.~\ref{fig: taylor vortex}(f), the green box (III) is significantly lower than the blue box (II) with reconstruction error being around $\epsilon_{p}/P_0 \approx 10\%$. 
		However, the solution from this `blunt' regularization is slightly distorted as shown in a typical pressure field and the corresponding error (Fig.~\ref{fig: taylor vortex}(d) and (i), respectively).
		This is due to the fact that we have inadvertently adjusted both the contaminated data ($\tilde{f}$) and the error-free boundaries (${g}_n$). 
		Despite ensuring compatibility, the exact error-free Neumann boundary conditions are no longer accurate and as a result can significantly affect pressure reconstruction \citep{faiella2021error,sperotto2022meshless}.
		
		In addition, one may notice that the error represented by the green box~(III) is higher than that of the red box~(I), despite that regularization with error-free Neumann boundary is utilized. 
		This suggests that a more careful approach is necessary, as when the contribution from the error on the boundary is absent, the total error in reconstructed pressure should be lower, especially for a small domain~\citep{pan2016error,pryce2024simple}.
		
		The approach proposed in test~IV is effective as shown in the orange box (IV) in Fig.~\ref{fig: taylor vortex}(f).
		The practice suggested by Note~\ref{note: div therom} grants the compatibility between $\tilde{f}_c$ and error-free $g_n$, while the reconstructed pressure field exactly satisfies the accurate Neumann conditions $g_n$. 
		The improved performance of this technique can also be seen in Fig.~\ref{fig: taylor vortex}(e\&j), where the pattern of the error is similar to that in the test~I (see Fig.~\ref{fig: taylor vortex}(b\&g). 
		The statistics of the reconstruction error show the expected error for test IV is lower than that for test I and test IV exhibits significantly lower variance (see Table~\ref{tab: mean and std 4 tests}).
		This recovers the intuition that proper use of accurate data can improve the accuracy and precision of the reconstruction.
		
		With the sample mean ($\mathbb{E}[||\epsilon_p||_{L^2(\Omega)}]$) and standard deviation ($\sigma_{\epsilon_p}$) of the error in the reconstruction, we may estimate the upper bound of the error $||\epsilon_p||_{L^2(\Omega)}$ using the 3-sigma rule.
		That is $||\epsilon_p||_{L^2(\Omega)} \lesssim \mathbb{U}[||\epsilon_p||_{L^2(\Omega)}]$, where $\mathbb{U}[||\epsilon_p||_{L^2(\Omega)}] = \mathbb{E}[||\epsilon_p||_{L^2(\Omega)}] + 3 \sigma_{\epsilon_p}$ is an estimated upper bound of the reconstruction error, which is presumably an uncertainty estimate. 
		In Table~\ref{tab: mean and std 4 tests}, we also list the upper bound estimates for the four tests. 
		It is evident that test~IV results in the lowest error due to compatible data and accurate boundary conditions, and test~II suffers from high error caused by the singularity at the Dirichlet corner due to incompatible underlying Neumann problem.


		\begin{table}[thbp!]
			\centering
			\caption{The sample mean, standard deviation, and the estimate of the upper bound of the reconstruction error in Fig.~\ref{fig: taylor vortex}(f).}
			\begin{tabular}{@{}ccccc@{}}
				\toprule
				& Test I & Test II & Test III & Test IV \\ 
				\midrule
				$\mathbb{E}[||\epsilon_p||_{L^2(\Omega)}]/P_0$ & $1.47 \times 10^{-2}$  & $4.42 $ & $1.04 \times 10^{-1}$  &    $1.41 \times 10^{-2}$  \\
				$\sigma_{\epsilon_p}/P_0$ & $1.21 \times 10^{-3}$  & $3.14 \times 10^{-1}$  & $3.49 \times 10^{-2}$   &  $5.51 \times 10^{-4}$  \\ 
				$\mathbb{U}[||\epsilon_p||_{L^2(\Omega)}]/P_0$ & $1.83 \times 10^{-2}$  & $5.36$  & $2.09 \times 10^{-1}$   &  $1.58 \times 10^{-2}$  \\
				\bottomrule
			\end{tabular}
			\label{tab: mean and std 4 tests}
		\end{table}

		\subsection{On the Iterative Solvers}
		\label{sec: On the iterative solvers}
		
\begin{figure}[htb]
			\centering
			\includegraphics[width=0.95\textwidth]{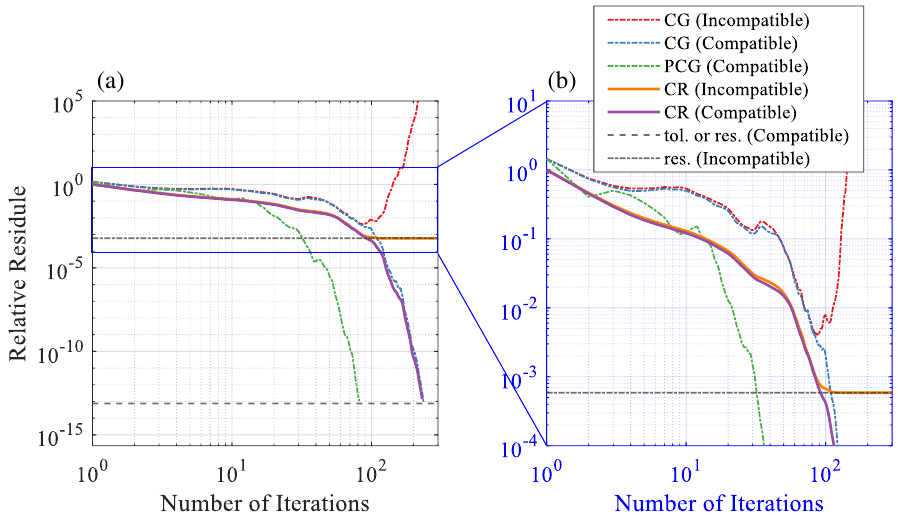}%
			\caption{(a) Behavior of iterative solvers (CG, CR, and  Preconditioned CG (PCG) using ILU) on compatible and incompatible linear systems derived from the PPE. (b) Zoom-in of the designated part of (a).}
			\label{fig: cg}
		\end{figure}

		To showcase the behaviour of the conjugate gradient solver used by \citet{zigunov2024one} and demonstrate other numerical minutiae related to linear solvers, we perform an additional experimentation where we compare CG and CR subject to compatible and incompatible data.
		We also show how simple preconditioning can effectively reduce the number of iterations required for convergence.
		In this numerical experiment, we employ the same Taylor vortex used in the previous numerical test in Sect.~\ref{sec: ODI vs. PPE} with the same all Neumann domain and discretization. 
		The contaminated pressure gradient and the data are also generated in a similar manner.
		To construct a compatible problem we use the data $\tilde{f}$ computed from contaminated gradient $\tilde{\bm{g}}$. 
		For the incompatible case we use the error-free Neumann boundary conditions $g_n$ and the noisy data $\tilde{f}$, which is similar to the practice for test II in Sect.~\ref{sec: Compatibility of PPE}. 
		The PPE is then solved using CG and CR for both the compatible and incompatible data from an initial guess of $p=0$ with the maximum iteration of $n_{\text{itr}} = 41^2 = 1681$ matching the dimension of $\bm{\mathcal{L}}_N$ and the tolerance being $\text{tol} = 7.6\times10^{-14}$.
		\textcolor{black}{We choose this tolerance as it is the residual of the solution from the pseudo-inverse computed by the SVD, which is approaching the level of machine tolerance.}
		Lastly, we also tested the Incomplete LU (ILU) preconditioned CG (PCG) on compatible data. 
		The results of this experiment can be seen in Fig.~\ref{fig: cg}.
		
		Figure~\ref{fig: cg} depicts the relative residual ($||\bm{\mathcal{L}}_N\hat{\vb{p}}_k - \vb{b}||_2 / ||\vb{b}||_2$) of the solution $\hat{\vb{p}}_k$ over the number of iterations $k$.
		As seen in Fig.~\ref{fig: cg}, CR is unconditionally stable independent of the compatibility of the system (purple and orange solid lines). 
		This is a critically favourable property as the solution is assured to not blow up. Thus, we recommend to use CR as the iterative solver (see Note~\ref{note: use CR}) for robust computation, especially when the consistency of the system is not known.
		For a compatible system, CR can converge to a small tolerance (on the level of machine precision, indicated by the horizontal gray dash line).
		\textcolor{black}{For an incompatible system, the residual of the CR algorithm plateaus above the tolerance (horizontal gray dashed line) and converges exactly at the residual of the solution from Moore-Penrose pseudo-inverse computed by an SVD-based solver (see the horizontal gray chain line).}
		This is expected, as the `exact' solution to an incompatible system does not exist; instead, CR converges to the MNLS solution, which is the same as the solution from the pseudo-inverse \citep{lim2024conjugate}.

		The behaviour of CG, on the other hand, greatly depends on the compatibility of the system. 
		On a compatible but semi-definite system, CG is guaranteed to converge to the minimal norm solution givin an intial guess of zero (see the blue chain line in Fig.~\ref{fig: cg}), which is a well-established result \citep{kammerer1972convergence,hestenes2012conjugate,lim2024conjugate}.
		Preconditioning can accelerate the solution (see the green chain line in Fig.~\ref{fig: cg} and Note~\ref{note: use preconditioning}). 
		For an inconsistent and semi-definite system again starting from zero, there is no guarantee that CG will not diverge after several iterations (see \citet{lim2024conjugate} for analysis).
		This is evident by observing the red chain line in Fig.~\ref{fig: cg}. 
		For the particular problem in our case, CG diverges after about 60 iterations on the inconsistent system. 
		However, before divergence, CG on the inconsistent and semi-definite system tends \textit{towards} the minimal norm solution similar to its behaviour on a compatible system. 
		Thus, if one has to use CG on an inconsistent and semi-definite system, an early-stopping criteria is essential (see Note~\ref{note: early-stopping for CG}). 
		Otherwise, some regularization should be applied to make the problem compatible.
		

		
		\section{Concluding Remarks and Perspectives}
		\label{sec: conclusion}
		By recovering the results in  \citet{zigunov2024one} through careful numerical treatment, we show that the Rotating Parallel Ray Omnidirectional Integration (RPR-ODI) is equivalent to pursuing the Minimum Norm (MN) or Minimum Norm Least Squares (NMLS) solution to the Pressure Poisson Equation (PPE) with all Neumann boundary conditions. 
		In doing so, we hope to put an end to the PPE versus ODI debate and clear up the confusion surrounding why and when these methods perform well.
		By examining image velocimetry based pressure field reconstruction---a classic challenge in experimental fluid mechanics---through the perspectives of the Poisson equation's well-posedness, linear algebra, estimation, and optimization, we provide insights into the strengths and limitations of conventional ODI methods \citep{liu2020error}, matrix ODI methods \citep{zigunov2023fast,zigunov2024one}, and select PPE-based solvers \citep{charonko2010assessment,mcclure2017optimization,zhang2020using}.
		
		We demonstrate that the key to successfully reconstructing the pressure field using ODI/PPE based on image velocimetry data lies in either i) formulating a compatible underlying Neumann problem and ensuring the uniqueness (e.g., by \textcolor{black}{pursuing an MN solution}), or ii) solving the corresponding inconsistent and semi-definite system by pursuing the MNLS solution.
		Both approaches ensure a well-posed and robust reconstruction.
		This understanding may inspire new regularization and data assimilation techniques or improve existing ones to further enhance the quality of pressure field reconstruction.
		
		\textcolor{black}{
			We also provide a general guideline for robust pressure field reconstruction based on the Neumann problem of PPE. 
			There are various paths towards the same reconstruction. 
			These methods address the well-posedness of the Poisson equation differently, however, all ensure the existence and uniqueness of the solution.
		}

		In terms of computational efficiency,  \citet{zigunov2023fast,zigunov2024one} already dramatically reduced the large computational cost of the conventional ODI to the level of PPE. 
		However, we recognize that the computational performance demonstrated in \citet{zigunov2023fast, zigunov2024one} and in the current work has room for improvement, as fast Poisson solvers remain an active area of research despite extensive existing results.
		By establishing the equivalence between ODI and PPE in the current work, we can enhance the computational efficiency of ODI or its equivalent PPE by leveraging the extensive literature on fast solvers for elliptic equations and robust numerical linear solvers.

		Before closing, we want to emphasize on two fundamental limitations (and potential improvements) of the conventional or matrix ODI algorithms as well as their equivalent PPE-based pressure solvers.
		First, given the fact that the error in the particle image velocimetry is usually correlated, the uncertainty in the domain is spatially inhomogeneous, and the error in the pressure gradient is not necessarily mean-zero, the reconstruction from this family of solvers does not grant the best linear unbiased estimator, or any estimator that minimizing the variance (i.e., uncertainty). 
		Instead, the MNLS solution is sub-optimal from a statistical perspective, and improvements to ODI/PPE can be readily achieved by applying established results in statistics and classical estimation theory \color{black} (e.g., generalized least squares is perhaps the most straightforward choice and see \citep{zhang2020using} for one implementation). \color{black}
		Second, ODI is motivated by the intention of enforcing the curl-free property for a corrupted pressure gradient field; however, the equivalency between the ODI and PEE shows that the curl-free correction goal is not achieved. 
		Thus, a pressure solver that properly incorporates curl-free correction could yield better results than the ODI and equivalent PPE-based solvers.
		Some existing works (e.g., \citet{wang2016irrotation,mcclure2017instantaneous,wang2017weighted,mcclure2019generalized,lin2023Divergence,li2024error}) can be seen as an improvement at different levels from this regard.

		\color{black}
		As discussed above, the insights presented in this work open up several avenues for enhancing ODI (or PPE) methods. In addition to the strategies already outlined throughout the current work, some straightforward approaches, such as using a larger stencil, may also be effective. 
		While implementing higher-order integration schemes in conventional ODI methods can be challenging, which  introduces even more computational cost, applying larger integration stencils is straightforward for a PPE-based solver. 
		Our preliminary tests suggest that discretizing the Laplacian with a larger stencil \citep{zigunov2023fast, zigunovpan2024private} leads to noticeable improvements in accuracy, especially when the data contains significant random noise components.

		\color{black}
		
		For now, we conclude the first part of our work on this topic, which primarily establishes the equivalence between ODI and PPE. 
		Further studies on potential improvements will follow.
		
		
		\section{Appendix}
		\appendix
		\color{black}
		\section{Development of the stencil by generic interpolation}
		\label{sec: generic interpolation}

		The stencil used by \citet{zigunov2024one} can also be developed in the framework of generic interpolation. Following the notations in Sect.~\ref{sec: FVM derivation} and writing the pressure gradient at each interface with neighboring cells as in \eqref{eq: grad p num}, we can approximate the gradient at the central node by interpolation
		\begin{equation}
			\label{eq: grad p interpolation LHS}
			\nabla p_i \approx \sum_j  \frac{h_{i,j} |\partial\Omega_i|_j}{\alpha|\Omega_i|}\frac{p_j - p_i}{h_{i,j}} \bm{n}_{i,j} =\sum_j  w_{i,j}\frac{p_j - p_i}{h_{i,j}} \bm{n}_{i,j} = \sum_j  w_{i,j} \nabla p_{i,j},
		\end{equation}
		where $\alpha = 4$ (or $6$) for two (or three) dimensional domains.
		$w_{i,j} = \frac{h_{i,j} |\partial\Omega_i|_j}{\alpha|\Omega_i|}$ can be viewed as weights, and \eqref{eq: grad p interpolation LHS} is a weighted average similar to the Sibson interpolation (also called area-stealing interpolation \citep{belikov2000non}).
		The `stolen areas' in this context are the triangles partitioning the central cell (e.g., shaded in purple with a vertex at the central node in Fig.~\ref{fig: voroni}, and the area of this triangle (or the volume of the tetrahedron) is $h_{i,j}|\partial\Omega_i|_j/\alpha$).
		In fact, as long as the weights form a partition of unity (i.e., $\sum_j w_{i,j} = 1$) and are non-negative (i.e., $w_{i,j} \geq 0$), they grant a valid and bounded interpolation, which is the case for \eqref{eq: grad p interpolation LHS}.
		For example, it is easy to verify that $w_{i,j}$ is the $j$-th partition of the total area (or volume) of the cell $\Omega_i$. 
		
		Invoking \eqref{eq: g = 0.5(g+g)} and using the same interpolation method above again to approximate the pressure gradient at the central node leads to
		\begin{equation}
			\label{eq: grad p interpolation RHS}
			\nabla p_i \approx \sum_j  \frac{h_{i,j} |\partial\Omega_i|_j}{\alpha|\Omega_i|}\frac{\bm{g}_i + \bm{g}_j}{2} = \sum_j  w_{i,j}\frac{\bm{g}_i + \bm{g}_j}{2}. 
		\end{equation}
		Equating \eqref{eq: grad p interpolation LHS} and \eqref{eq: grad p interpolation RHS} recovers \eqref{eq: sum p-p = sum g+g}.
		
		A similar derivation by the partition of unity also holds for cells adjacent to boundaries.
		We do this by introducing a ghost cell $\Omega_k$ outside the boundary of the domain that mirrors the node $i$. 
		This means that the interface between cells $\Omega
		_{i,k}$ (see the thick purple line in Fig.~\ref{fig: voroni}(a)) is in the middle of the nodes $i$ and $k$ and is perpendicular to the line connecting the two nodes. 
		The measured pressure gradient outside the domain can be estimated by linear extrapolation $\bm{g}_k \approx
		\bm{g}_i$ and the pressure at the ghost node can be approximated by 
		\begin{equation}
			\label{eq: OSMODI BC}
			p_k \approx  p_i + h_{i,k} \bm{g}_i \cdot\bm{n}_{i,k},
		\end{equation}
		which is rearranged from  $\nabla p_i \approx \frac{p_k - p_i}{h_{i,k}} \bm{n}_{i,k} \approx \bm{g}_i$, meaning that  \eqref{eq: OSMODI BC} implements a Neumann boundary condition (i.e., $\nabla p \cdot \bm{n} = g_n$ on $\partial \Omega$). 
		
		Substituting the above approximations for $\bm{g}_k$ and $p_k$ into \eqref{eq: sum p-p = sum g+g}, we can see that the contributions from the ghost cell and node cancel out, which is equivalent to setting a vanishing weight for the node outside the domain (i.e., $w_{i,k} = 0$). With this boundary treatment, \eqref{eq: pc = sum pj - g+g} is recovered.
		This treatment easily generalizes to multiple ghost cells outside the domain, which may be useful for boundaries with sharp corners.
		In \citet{zigunov2024one}, they implemented this boundary treatment using boolean numbers.

		\section{on the curl-free correction of ODI/PPE}
		\label{sec: appendix: curl-free correction}
		
		\textcolor{black}{
			Constructing a Poisson equation by $\nabla^2 \tilde{p} = f = \nabla \cdot \nabla \tilde{\bm{g}}$ performs curl-free construction to the corrupted data $\tilde{\bm{g}}$ in the domain but not on the boundary. 
			Unfortunately, the (solenoidal) errors on the boundaries are influential; and solving a such Poisson equation alone using MN or MNLS solvers does not fix the problem.
			This argument is evident by applying Helmholtz-Hodge Decomposition (HHD) to $\tilde{\bm{g}}$.
			One version of HHD \citep{bhatia2012helmholtz} states that the contaminated pressure gradient measurement can be separated into two components: 
			\begin{equation}
				\label{eq: HHD}
				\tilde{\bm{g}} = \nabla \phi + \nabla \times \bm{r},
			\end{equation}
			where $\phi \approx {p}$ is an approximate of the pressure, which is a scalar field, and thus $\nabla {\phi}$ is curl-free.
			$\nabla \times  \bm{r}$ is the solenoidal component and should not be a part of the pressure gradient.
			Taking divergence of \eqref{eq: HHD} leads to a Poisson equation:
			\begin{equation}
				\label{eq: HHD Poisson}
				\nabla^2 p \approx \nabla^2 \phi = \nabla \cdot \tilde{\bm{g}} = \tilde{f},
			\end{equation}    
			where the divergence-free component $\nabla \times  \bm{r}$ is removed by invoking the identity $\nabla \cdot (\nabla \times  \bm{r}) = 0$.}
		
		However, $\tilde{\bm{g}}$ on the boundary remains uncorrected.
		Specifically, the solenoidal components in $\tilde{\bm{g}}$ on the boundary would result in erroneous Neumann boundary conditions (i.e., $\nabla \phi \cdot \bm{n} =  \tilde{\bm{g}}\cdot \bm{n} = \tilde{g}_n$), which could significantly corrupt the solution to \eqref{eq: HHD Poisson}.
		In other words, establishing the Poisson equation as in \eqref{eq: HHD Poisson}, which is what ODI methods implicitly do (see Sect.~\ref{sec: ODI = PPE}), fails to correct the solenoidal/erroneous components on the boundary---which is critical---the same way as common implementation of PPE.
		This motivates the use of HHD for better pressure field reconstruction.

		\section{on the nuances of LS, MN, and MNLS solutions}
		\label{sec: appendix: LS, MN, and MNLS solutions}
		
		
		Here we provide a brief overview of solutions for the general linear system $ \vb{A}\vb{x} = \vb{b} $, where $ \vb{A} \in \mathbb{R}^{m \times n} $, $ \vb{x} \in \mathbb{R}^n $, and $ \vb{b} \in \mathbb{R}^m $ in the context of the current work. 
		This appendix serves as a reference for housekeeping and readers' convenience. 
		The following content can be found in standard linear algebra textbooks.
		
		\subsection{Injectivity and Surjectivity}  
		
		A matrix $ \vb{A} \in \mathbb{R}^{m \times n} $ is surjective (onto) if and only if its rank equals the number of rows, i.e., $ r = \text{rank}(\vb{A}) = m $. 
		Surjectivity ensures that $ \vb{A}\vb{x} $ spans all of $ \mathbb{R}^m $. 
		$ \vb{A} \in \mathbb{R}^{m \times n} $ is injective (one-to-one) if and only if its rank equals the number of columns, i.e., $  r = \text{rank}(\vb{A}) = n $. 
		Injectivity ensures that $ \vb{A}\vb{x} $ is unique for any $ \vb{x} \in \mathbb{R}^n $. 
		A square matrix of full rank is both onto and one-to-one; a tall matrix $(m>n)$ of full rank is only one-to-one; a fat matrix $(m<n)$ of full rank is only onto; and lastly, any sized matrix that is rank deficient is neither onto nor one-to-one. 
		When $\vb{A}$ is onto and one-to-one (square and full rank), the solution to $\vb{A}\vb{x} = \vb{b}$ can be found using the standard matrix inverse $\vb{x} = \vb{A}^{-1}\vb{b}$, which always exist and is unique. 
		However, this is not always the case, and a pseudo-inverse may be needed, which is summarized as follows.

		\subsection{Least Squares Solution}
		
		The standard way to solve overdetermined $(m>n)$ systems of linear equations is to look for the Least Squares (LS) solution.
		More specifically, if $\vb{A}$ is not onto, $\vb{b}\notin \text{Col}(\vb{A})$ and there is no $\vb{x}$ that perfectly satisfies $ \vb{A}\vb{x} = \vb{b}$.
		In this case, we instead look to find the solution that best satisfies $\vb{A}\vb{x} = \vb{b}$ by solving the following optimization problem:
		\begin{align}
			\label{eq: least squares}
			\min_{\vb{x}} ~   & \mathcal{J} = \|\vb{A} \vb{x} - \vb{b}\|_2^2.
		\end{align}
		The solution to \eqref{eq: least squares} always exists, however, it is unique if $\vb{A}$ is one-to-one.
		If $\vb{A}$ is not one-to-one then there are infinitely many solutions that have the same minimum residual. 
		Assuming $\vb{A}$ is one-to-one, an explicit solution to \eqref{eq: least squares} is
		\begin{equation}
			\label{eq: least squares solution}
			\hat{\vb{x}} = \vb{A}^\dagger \vb{b} = ( \vb{A}^\intercal\vb{A})^{-1}\vb{A}^\intercal\vb{b},
		\end{equation}
		which is known as the solution to the left normal equation, where $ \vb{A}^\dagger = ( \vb{A}^\intercal\vb{A})^{-1}\vb{A}^\intercal$ is the left pseudo inverse.
		The fundamental problem with an over-determined system is the lack of existence of a solution, meaning there might not be any linear combination of the columns of $\vb{A}$ that can produce the data $\vb{b}$. 
		LS addresses this issue by projecting $\vb{b}$ onto the column space of $\vb{A}$. 
		
		
		\subsection{Minimum Norm Solution}
		
		A typical solution to an underdetermined $(m<n)$ system is to look for the Minimum Norm (MN) solution. 
		Formally, $\vb{A}$ is not one-to-one and the solution is not unique. 
		However, the lack of uniqueness can be resolved by introducing additional constraints or assumptions. 
		For example, one can pursue the 
		smallest solution that satisfies the linear system. 
		This idea can also be formulated as an optimization problem:
		\begin{align}
			\label{eq: minimum norm}
			\begin{split}
				\min_{\vb{x}} ~   & \mathcal{J} = \|\vb{x} \|_2^2 \\
				\text{s.t.} ~ & \vb{A} \vb{x} = \vb{b}.
			\end{split}
		\end{align}
		The solution to \eqref{eq: minimum norm} is always unique, however, it only exists if $\vb{A}$ is onto. 
		Assuming $\vb{A}$ is onto, the analytical solution to \eqref{eq: minimum norm} is 
		\begin{equation}
			\label{eq: minimum norm solution}
			\hat{\vb{x}} = \vb{A}^{\dagger} \vb{b} = \vb{A}^\intercal(\vb{A}\vb{A}^\intercal)^{-1}\vb{b},
		\end{equation}
		which is known as the solution to the right normal equation, and $\vb{A}^{\dagger} = \vb{A}^\intercal(\vb{A}\vb{A}^\intercal)^{-1}$ is the right pseudo inverse.
		
		
		The fundamental problem with an under-determined system is the lack of uniqueness. 
		In this case, the solution can always be written as $\vb{x} = \vb{x}_{\text{p}}+\vb{x}_{\text{null}},$ where $\vb{x}_{\text{p}}$ is a particular solution so that $\vb{A}\vb{x}_{\text{p}} = \vb{b}$, and  $\vb{A}\vb{x}_{\text{null}} = \vb{0}$, and $\vb{x}_{\text{null}}$ is a general solution.
		After satisfying the linear system with $\vb{x}_{\text{p}}$ there are still degrees of freedom to vary the solution by adding $\vb{x}_{\text{null}}$ to $\vb{x}_{\text{p}}$.
		By asking for the MN solution, we look for the smallest solution that satisfies the linear system which forces $\vb{x}_{\text{null}} = \vb{0}$.

		Next, we investigate a situation raised from the context in the current work, which is slightly more nuanced. 
		In general, $\vb{A}$ must be onto for the minimal norm problem to be well defined; however, this requirement is a sufficient condition.
		The necessary condition instead states that $\vb{b}\in\text{Col}(\vb{A})$. 
		This point is especially important in cases where $\vb{A}$ is not onto and $\vb{b}$ is not compatible with $\vb{A}$.
		A good example of this situation is the discrete Laplacian matrix $\bm{\mathcal{L}}_N$ discussed in the current work which is square and rank deficient. 
		The matrix $\bm{\mathcal{L}}_N$ may not be onto, however, an MN solution exists if $\vb{b}$ is compatible. 
		If $\vb{b}_N$ is perturbed, or calculated inconsistently, the problem may no longer be well-defined and a MN solution does not exist. 
		In this case, one may pursue Minimum Norm Least Squares (MNLS) Solutions instead. 

		\subsection{Minimum Norm Least Squares Solutions}
		
		
		To solve rank deficient problems where $\vb{A}$ is neither onto nor one-to-one, one can pursue a Minimum Norm Least Squares (MNLS) solution to the system by the Moore-Penrose pseudo-inverse.
		
		
		One generic way to define the Moore-Penrose pseudo-inverse is by the Singular Value Decomposition (SVD), which factorizes $\vb{A}$ of rank $q$ into the product of three matrices $\bm{U}\in\mathbb{R}^{m\times m}$, $\bm{S}\in\mathbb{R}^{m\times n}$, and $\bm{V}\in\mathbb{R}^{n\times n}$, such that $\vb{A} = \vb{U}\vb{S}\vb{V}^\intercal.$
		The first $q$ columns of $\vb{U}$ and $\vb{V}$ form an orthonormal basis for the column and row space of $\vb{A}$ respectively.
		$\vb{S}$ is diagonal and contains the singular values of $\vb{A}$ ordered from largest to smallest. 
		The Moore-Penrose pseudo inverse is defined in terms of the SVD as $\vb{A}^\dagger = \bm{V}\bm{S}^\dagger\bm{U}^\intercal,$ where $\bm{S}^\dagger$ is calculated by taking the reciprocal of all of the non-zero singular values. 
		Given the Moore-Penrose pseudo inverse, the solution to an arbitrary linear system can be achieved as follows:
		\begin{equation}
			\label{eq: pseudoinverse solution}
			\hat{\vb{x}} = \vb{A}^\dagger \vb{b}  = \bm{V}\bm{S}^\dagger\bm{U}^\intercal \vb{b}.
		\end{equation}

		The Moore-Penrose pseudo-inverse defined above implicitly changes its behaviour depending on the properties of $\vb{A}$ to work for any $\vb{A}$.
		If $\vb{A}$ is onto and one-to-one (square and full rank), $\vb{A}^\dagger$ recovers the standard matrix inverse. 
		If $\vb{A}$ is only one-to-one, \eqref{eq: pseudoinverse solution} provides the least squares solution. 
		Lastly, if $\vb{A}$ is only onto, \eqref{eq: pseudoinverse solution} results in the minimum norm solution.

		The Moore-Penrose pseudo inverse also extends to problems where $\vb{A}$ is neither onto nor one-to-one by
		treating the incompatibility in $\vb{A}$ using least squares and the non-uniqueness in $\vb{A}$ with minimum norm simultaneously. 
		To demonstrate how this works we can rewrite \eqref{eq: pseudoinverse solution} as a sum of vector products
		\begin{equation}
			\label{eq: pseudoinverse solution sum}
			\hat{\vb{x}} = \sum_{i=1}^q \frac{1}{s_i}\vb{u}_i^\intercal\vb{b}\vb{v}_i.
		\end{equation}
		Here the solution is expressed as the linear combination of $\vb{v}_i$, the first $q$ column vectors of $\bm{V}$, which are weighted by the projection of $\vb{b}$ onto the first $q$ columns of $\bm{U}$, $\vb{u}_i$, divided by the non-zero singular values $s_i$. 
		
		It is evident from \eqref{eq: pseudoinverse solution sum} how the Moore-Penrose pseudo inverse achieves both the LS and MN solution at the same time: i)  
		since the first $q$ columns of $\bm{U}$ form a basis for the column space of $\vb{A}$, $\vb{u}_i^\intercal\vb{b}$ projects $\vb{b}$ onto the column space of $\vb{A}$ like least squares. 
		Similarly, the first $q$ columns of $\bm{V}$ form a basis for the row space, meaning $\vb{x}$ has no terms that correspond to the null space of $\vb{A}$ and thus the solution is of minimum norm. 
		\textcolor{black}{We also note that some sources call the MNLS solution simply the least squares solution, however, in the current work we wish to clarify the nuances.}

		
		\section{Practical computation for MNLS}
		\label{sec: appendix: Tikonov and CG for SVD}
		
		SVD is a powerful tool of theoretical interest, however, in practice, full singular value decompositions for large matrices can be expensive.
		For this reason, obtaining an approximate MNLS solution at a lower cost is attractive in practice. 
		One simple way to do this is by the Tikhonov regularization. 
		In this framework, the solution minimizing $\mathcal{J} = \|\vb{A}\vb{x}-\vb{b}\|_2^2+\lambda^2\|\vb{x}\|_2^2$ always exists and is unique for a nonzero $\lambda$.
		This problem has a simple closed-form solution $$\hat{\vb{x}} = \sum_{i=1}^q \frac{s_i}{s_i^2 + \lambda^2} \vb{u}_i^\intercal\vb{b}\vb{v}_i$$ and can be achieved fast using iterative schemes by solving $(\vb{A}^\intercal\vb{A}+\lambda\vb{I})\vb{x}=\vb{A}^\intercal\vb{b}$.
		As $\lambda$ gets small, the Tikhonov solution approximates the pseudo inverse solution.
		When choosing some larger $\lambda$, Tikhonov regularization can effectively improve the conditioning of the original problem, which may be useful to reduce the sensitivity of the reconstruction against perturbations to the data (see Sect.~\ref{sec: Continuous Dependence}).
		
		
		Other techniques include iterative regularization methods such as CG with early stopping criteria and an initial guess of zero solution. 
		Each CG step is analogous to a term in \eqref{eq: pseudoinverse solution sum} and varies the solution from $\hat{\vb{x}}=\bm{0}$, which is the smallest possible solution. This is an alternative way to interpret the success of the \cite{zigunov2024one}.
		Tikhonov, CG-like regularization, and practical computation of MNLS are entire topics of their own and will be saved for future discussions.

		\color{black}

		\section*{Acknowledgement}
		The authors thank Dr. Fernando Zigunov for the discussions. 

		\color{black}
		
		
%
		
%
		
		\subsection*{Authors' contributions}
		Conceptualization: Z.P.; formal analysis:  Z.P. and C.P.; programming: C.P., Z.P., and L.L.; writing and editing: Z.P., C.P., and L.L.

		\subsection*{Funding}
		This work is partially supported by the Natural Sciences and Engineering Research Council of Canada (NSERC) Discovery Grant (RGPIN-2020-04486),  Undergraduate Research Assistantship (URA), and Undergraduate Student Research Awards (USRA) programs at University of Waterloo.

		\subsection*{Availability of data and materials}
		Datasets and codes are available from the corresponding author, Z.P. upon reasonable request.

		\bibliographystyle{apalike}
		\bibliography{PIV_Pressure_Lib}
	\end{document}